\documentclass[a4paper,12pt]{article}
\usepackage{mathrsfs}
\usepackage{amsmath}
\usepackage{amscd}
\usepackage{amssymb}
\usepackage{cite}
\usepackage[dvips]{color}
\renewcommand{\theequation}{\arabic {section}.\arabic{equation}}

\allowdisplaybreaks

\begin{document}


\begin{titlepage}
\begin{flushright}
EPHOU-08-004 \\
August, 2008
\end{flushright}

\vspace{5cm}

\begin{center}
{\Large Vafa-Witten Theory on $N$=2 and $N$=4 Twisted Superspace in Four
 Dimensions} \\ 

\vspace{1cm}

{\scshape Junji Kato}\footnote{jkato@particle.sci.hokudai.ac.jp}, 
{\scshape Akiko Miyake}\footnote{miyake@particle.sci.hokudai.ac.jp}\\

{\textit{ Department of Physics, Hokkaido University }}\\
{\textit{ Sapporo, 060-0810, Japan}}\\
\end{center}

\vspace{2cm}

\begin{abstract}
We construct a new off-shell twisted hypermultiplet with a scalar and
an anti-self-dual tensor superfields. 
Using the $N=2$ twisted superspace formalism, we construct a
 Donaldson-Witten theory coupled to the hypermultiplet. We show that
 this action possesses the Vafa-Witten type $N=4$ twisted supersymmetry 
at the on-shell level.
We also reconstruct the action
 using a $N=4$ twisted superconnection formalism.
\end{abstract}

\end{titlepage}

\newpage
\renewcommand{\theequation}{\arabic {section}.\arabic{equation}}

\section{Introduction}
 
One of the important characteristic of a topological twist \cite{W1,bs} 
is that the BRST charge of a quantized topological field theory
relates to a supercharge. One can construct a new algebra which is called
twisted superalgebra by using the twisting procedure. These algebras
consist of scalar, vector and tensor generators. The twisted
supersymmetry is investigated in various models and various
dimensions \cite{BR1,BR2,BR3,BRT,DGS,GMS1,GMS2,BMa,FTVVSS,LSZ,GGPS,CLS,LL,LSSTV}. We are especially interested in Dirac-K\"ahler twist
\cite{KT} which is connected with lattice fermion. 
The advantage of this twist is that ghost fields in the quantized
topological field theory are directly related to the gaugino or matter
fields in supersymmetric theory and that an untwisted theory is easily
constructed from the twisted supersymmetric theory. A twisted
superspace formalism are then constructed from the twisted superalgebra
\cite{BT,AL}. 
Twisted supersymmetric Yang-Mills theories on the superspace formulation
based on the Dirac-K\"ahler twist was derived in two and four dimensions
\cite{KKU,KKM}. Three dimensional twisted supersymmetric Yang-Mills
theory was investigated in \cite{N1}. 
A recent development of the twisted superspace is that a path-integral
quantization procedure with respect to the subsuperspace which consists
of scalar and vector fermionic coordinates is proposed in four dimensional
$N$=2 super Yang-Mills case \cite{BBM,BM}.

In previous paper\cite{KM} we constructed a $N$=2 twisted superspace
formalism with a central charge using the Dirac-K\"ahler twist. 
Our formulation was based on a tensor formulation coming from the
Dirac-K\"ahler twisting procedure. We proposed a new $N$=2 twisted
hypermultiplet with a central charge. This multiplet includes a bosonic
vector field. We then proposed a new
off-shell twisted hypermultiplet action and gave a gauge covariant
version of this action. It turned out that this action plus
Donaldson-Witten action has the on-shell $N$=4 TSUSY and the
four-dimensional Dirac-K\"ahler twist is equivalent to the Marcus's
twist \cite{Mac}.  

In this paper we propose a new $N$=2 twisted hypermultiplet with a
central charge. This multiplet consist of a bosonic scalar and a bosonic
self-dual antisymmetric tensor fields. In this case a covariantized
on-shell 
action similarly possesses a $N$=4 twisted supersymmetry. An algebra of
this symmetry is different from the Marcus's one. An assignment of ghost
number is especially different in these theories. We insist that this
theory is equivalent to the Vafa-Witten theory\cite{Yam,VW}.  

Another important motivation of this work comes from the recent study of
lattice SUSY. It is well known that the Dirac-K\"ahler fermion mechanism
is well-defined on the lattice\cite{KSuss,Suss,KS}. Supersymmetric
models with modified Leibniz rule on the lattice was studied in
\cite{DKKN1-1,DKKN1-2,DKKN2, NW}. The characteristic of these models is that
all of the twisted supersymmetries are exactly defined on the lattice. 
Other models by using the Dirac-K\"ahler fermion without modified
Leibniz rule possess the partial twisted supersymmetries. These models
was investigated in \cite{C2, C1-1, C1-2, C5, C6, C7, S1, S2, S3, S4, S5}. 
In the recent development of lattice SUSY, there are matrix formulations
on which we impose $Z_N$ orbifold conditions\cite{KKU1, CKKU1, CKKU2, KU,
Kaplan, Unsal1, Unsal2,
Onogi-T, Ohta-T, Takimi, DM1, DM2, DM3, Matsuura1, DM4, Matsuura2, ADFKS}.
These models are constructed
 under the orbifold condition coming from some global symmetries of some
 mother theories. The lattice SUSY models based on the twisted SUSY may
 be related to the one based on the matrix formulation.

This paper is organized as follows. In Sec. 2 we give a brief
introduction of the $N$=2 twisted superspace formulation with a central
charge based on the Dirac-K\"ahler twist. In Sec. 3 we propose a new
twisted hypermultiplet which is constructed by the scalar and tensor
superfields. In Sec. 4 we derive a Donaldson-Witten theory coupling to
the hypermultiplet. We show that the supersymmetries of the theory become
the $N$=4 twisted supersymmetry which is correspondence with Vafa-Witten
type one. In Sec. 5  we reconstruct the $N$=4 twisted SYM theory with
respect to the $N$=4 twisted superspace formulation. We summarize the
results in section 5. We provide several appendices to summarize the
notations and show the full transformation of on-shell $N$=4 TSUSY.

\section{$N=2$ twisted SUSY with central charge}
In a previous paper \cite{KM} we derived the N=2 twisted SUSY algebra and
superspace formalism with a central charge with respect to the
Dirac-K\"ahler fermion \cite{IL,Kahler}. This algebra is a twisted version of the ordinary
$N$=2 SUSY algebra with a central charge \cite{Fay,Soh}.  
$N=2$ twisted SUSY generators consist of a scalar, vector,
anti-self-dual tensor and central charge $\{s^+, s^+_\mu,s^+_{\mu\nu}
\ \mbox{and}\ Z \}$. The algebra of the twisted supercharges are 
\begin{align}
\{ s^+, s^+_\mu\} &= P_\mu ,& \{ s^+ _A , s^+_\mu \} &= -\delta^+
 _{A,\mu\nu} P^\nu,& \{s^+, s^+_A\} &= 0, \nonumber \\
\{s^+, s^+\} &= Z ,& \{s^+_\mu,s^+_\nu\} &= \delta_{\mu\nu} Z ,&
 \{s^+_A, s^+_B\} &=  \delta^+ _{A,B} Z,
\label{eq:TSUSY al. with central charge}
\end{align}
where the others (anti)commute,
the capital $\{A\}$ denotes the second rank tensor indices $\mu\nu$ and
$\delta^+_{\mu\nu,\rho\sigma}$ ($\delta^+_{A,B}$) is defined as
$\delta^+_{\mu\nu,\rho\sigma}\equiv\delta_{\mu\rho}
\delta_{\nu\sigma}-\delta_{\mu\sigma}\delta_{\nu\rho}-
\epsilon_{\mu\nu\rho\sigma}$.
Through this paper we consider the Euclidean flat spacetime.

We now introduce the $N=2$ twisted superspace based on the
algebras. The superspace consist of the 
bosonic coordinates $x_\mu$, $z$ and
fermionic ones $\theta^+$, $\theta^+_\mu$, $\theta^+_{\mu\nu}$, where
$z$ is a bosonic parameter corresponding to the central charge $Z$ and
$\theta^+_{\mu\nu}$ are fermionic anti-self-dual tensor parameters
with $\theta^+_{\mu\nu}=-\frac{1}{2}\epsilon_{\mu\nu\rho\sigma}
\theta^{+\rho\sigma}$. We define differential operators
$\{\mathcal{Q}^+, \mathcal{Q}^+_\mu, \mathcal{Q}^+_{A}, \mathcal{Z}\}$
corresponding with the twisted supercharges $\{s^+, s^+_\mu, s^+_{A}, Z\}$
as follows:
\begin{eqnarray}
\mathcal{Q}^+ &=& \frac{\partial}{\partial \theta^+} +\frac{i}{2}
 \theta^{+\mu}\partial_\mu +\frac{i}{2}\theta^+ \frac{\partial}{\partial
 z}\nonumber, \\
\mathcal{Q}^+_{\mu} &=& \frac{\partial}{\partial \theta^{+\mu}}
 +\frac{i}{2} \theta^{+}\partial_\mu-\frac{i}{2}\theta^+
 _{\mu\nu}\partial^\nu +\frac{i}{2}\theta^+_\mu \frac{\partial}{\partial
 z}, \nonumber \\
\mathcal{Q}^+_A &=& \frac{\partial}{\partial \theta^{+A}}
-\frac{i}{2} \delta^+ _{A,\mu\nu}\theta^{+\mu}\partial^\nu
 +\frac{i}{2}\theta^+_A \frac{\partial}{\partial z}, \nonumber \\
\mathcal{Z}&=& -i\frac{\partial}{\partial z}.
\end{eqnarray}
These differential operators satisfy the following anticommuting relations:
\begin{align}
\{ \mathcal{Q}^+,\mathcal{Q}^+ _\mu\} &= i\partial_\mu ,&
\{ \mathcal{Q}^+ _A , \mathcal{Q}^+ _\mu \} &= -i\delta^+ _{A,\mu\nu}
 \partial^\nu,& \{\mathcal{Q}^+,\mathcal{Q}^+_A\}&=0, \nonumber \\
\{\mathcal{Q}^+,\mathcal{Q}^+\} &=- \mathcal{Z} ,&
 \{\mathcal{Q}^+_\mu,\mathcal{Q}^+_\nu\}& =-\delta_{\mu\nu} \mathcal{Z} ,&
 \{\mathcal{Q}^+_A,\mathcal{Q}^+_B\} &= - \delta^+ _{A,B} \mathcal{Z},
\end{align}
where $\frac{\partial}{\partial \theta^{+A}}\theta^+_B\equiv
\delta^+_{A,B}$ and $\mathcal{Z}$ commutes all the differential
operators. The sign of the spacetime
derivative is reversed with respect to the algebra (\ref{eq:TSUSY
al. with central charge}). We then introduce differential operators
$\{\mathcal{D}^+, \mathcal{D}^+_\mu, \mathcal{D}^+_A\}$ which
anticommute with the differential operators $\{\mathcal{Q}^+,
\mathcal{Q}^+_\mu, \mathcal{Q}^+_{A}\}$,
\begin{eqnarray}
\mathcal{D}^+ &=& \frac{\partial}{\partial \theta^+} -\frac{i}{2}
 \theta^{+\mu}\partial_\mu -\frac{i}{2}\theta^+ \frac{\partial}{\partial
 z} \nonumber, \\
\mathcal{D}^+ _{\mu} &=& \frac{\partial}{\partial \theta^{+\mu}}
 -\frac{i}{2} \theta^{+}\partial_\mu + \frac{i}{2}\theta^+ _{\mu\nu}
 \partial^\nu -\frac{i}{2}\theta^+_\mu \frac{\partial}{\partial z},
 \nonumber \\
\mathcal{D}^+ _A &=& \frac{\partial}{\partial \theta^{+A}}
 +\frac{i}{2} \delta^+ _{A,\rho\sigma}\theta^{+\rho}\partial^\sigma
 -\frac{i}{2}\theta^+_A \frac{\partial}{\partial z}.
\end{eqnarray}
These operators $\{\mathcal{D}^+, \mathcal{D}^+_\mu, \mathcal{D}^+_A\}$
satisfy the following relations:
\begin{align}
\{ \mathcal{D}^+,\mathcal{D}^+ _\mu\} &= -i\partial_\mu ,&
\{ \mathcal{D}^+ _A , \mathcal{D}^+ _\mu \} &= i\delta^+ _{A,\mu\nu}
 \partial^\nu,& \{\mathcal{D}^+,\mathcal{D}^+_A\}&=0, \nonumber \\
\{\mathcal{D}^+,\mathcal{D}^+\} &= \mathcal{Z} ,&
 \{\mathcal{D}^+_\mu,\mathcal{D}^+_\nu\}& =\delta_{\mu\nu} \mathcal{Z} ,&
 \{\mathcal{D}^+_A,\mathcal{D}^+_B\} &=  \delta^+ _{A,B} \mathcal{Z},
\end{align}
where $\mathcal{Z}$ commutes these operators $\{\mathcal{D}^+,
\mathcal{D}^+_\mu, \mathcal{D}^+_A\}$.

\section{A new twisted hypermultiplet}
In previous paper we constructed the twisted hypermultiplet introducing
the bosonic superfield with vector index. We now introduce bosonic
superfields with a scalar and an anti-self-dual tensor indices,
$\mathcal{V}^+$ and $\mathcal{V}^+_A$ respectively. 
Since a general superfield has many component fields,
we need to eliminate superfluous fields.
We then use the R-symmetry in order to impose a condition on the
superfield.   
We also introduce R-transformations for the superfield $\mathcal{V}^+,\mathcal{V}^+_A$
and supercharges:  
\begin{eqnarray}
R^+_A \mathcal{V}^+ &=&- \frac{i}{2} \mathcal{V}^+_A ,\nonumber \\
R^+_A  \mathcal{V}^+_B &=& \frac{i}{2} \delta^+ _{A,B}\ \mathcal{V}^+ -\frac{i}{8} \Gamma^+ _{ABC}\ \mathcal{V}^{+C} ,\nonumber \\
R^+_A s^+_\mu &=& -\frac{i}{2}\delta^+ _{A,\mu\nu} s^{+\nu} ,\nonumber \\
R^+_A s^+ &=&- \frac{i}{2} s^+_A ,\nonumber \\
R^+_A  s^+_B &=& \frac{i}{2} \delta^+ _{A,B}\ s^+ -\frac{i}{8} \Gamma^+ _{ABC}\ s^{+C} ,\nonumber \\
R^+_A s^+_\mu &=& -\frac{i}{2}\delta^+ _{A,\mu\nu} s^{+\nu} ,
\label{eq:R-symmetry}
\end{eqnarray}
where $\Gamma^+_{ABC}$ is an anti-symmetric tensor defined in the
Appendix. The $\{\mathcal{D}^+\}$ operators transform in the same
manner with respect to the supercharges. 
We can find the following R-invariant terms,
\begin{eqnarray}
&&R^+_A (\mathcal{D}^+_\mu \mathcal{V}^+ +\mathcal{D}^{+\nu}
 \mathcal{V}^+ _{\mu\nu})=0, \nonumber\\
&&R^+_A (\mathcal{D}^+\mathcal{V}^++\frac{1}{4}\mathcal{D}^{+B}
 \mathcal{V}^+_B)=0, \nonumber\\
&&R^+_A (\mathcal{D}^+_B \mathcal{V}^+-\mathcal{D}^+\mathcal{V}^{+}_B
 + \frac{1}{16}\Gamma^+_{BCD} \mathcal{D}^{+C} \mathcal{V}^{+D})=0.
 \label{eq:R-inv}
\end{eqnarray}
We may impose the following conditions:
\begin{eqnarray}
R^+_A(\mathcal{D}^+\mathcal{V}^+)=R^+_A(\mathcal{D}^+_\mu\mathcal{V}^+)
 =R^+_A(\mathcal{D}^+_B\mathcal{V}^+)=0, \nonumber\\
R^+_A(\mathcal{D}^+\mathcal{V}^+_B)=R^+_A(\mathcal{D}^+_\mu\mathcal{V}^+_B)
 =R^+_A(\mathcal{D}^+_B\mathcal{V}^+_C)=0. \label{eq:R-constraints}
\end{eqnarray}
The constraints (\ref{eq:R-constraints}) mean that $\mathcal{D}^+_I
\mathcal{V}^+$ and $\mathcal{D}^+_I \mathcal{V}^+_A$ should be the
R-invariant, where $\{\mathcal{D}^+_I\}=\{\mathcal{D}^+,\mathcal{D}^+_\mu,\mathcal{D}^+_A\}$.
We can then find the following relations between $\mathcal{V}^+$ and
$\mathcal{V}^+_A$ by using eqs. (\ref{eq:R-inv}) and
(\ref{eq:R-constraints}):
\begin{eqnarray}
& &\delta^+_{A,\mu\nu}\mathcal{D}^{+\nu} \mathcal{V}^+
 +\mathcal{D}^+_\mu \mathcal{V}^+ _A =0, \nonumber\\
& &\mathcal{D}^+_A \mathcal{V}^+ +\mathcal{D}^+ \mathcal{V}^+_A =0,
 \nonumber\\
& &\mathcal{D}^+_A \mathcal{V}^+_B +\mathcal{D}^+_B \mathcal{V}^+_A
 -2\delta^+_{A,B} \mathcal{D}^+\mathcal{V}^+ =0, \nonumber\\
& &\mathcal{D}^+_A \mathcal{V}^+_B - \mathcal{D}^+_B \mathcal{V}^+_A
 -\frac{1}{4}\Gamma^+_{ABC}(\mathcal{D}^{+C}\mathcal{V}^+-\mathcal{D}^+
 \mathcal{V}^{+C} )=0. \label{eq:constraints}
\end{eqnarray}
We then derive the following relations from the constraints (\ref{eq:constraints}):
\begin{eqnarray}
\mathcal{D}^+\mathcal{D}^+\mathcal{V}^+&=&\frac{1}{2}\mathcal{ZV}^+
 , \nonumber\\
\mathcal{D}^+\mathcal{D}^+_A\mathcal{V}^+&=&-\frac{1}{2}\mathcal{ZV}^+_A
 , \nonumber\\
\mathcal{D}^+_A\mathcal{D}^+\mathcal{V}^+&=&\frac{1}{2}\mathcal{ZV}^+_A
 , \nonumber\\
\mathcal{D}^+_A\mathcal{D}^+_B\mathcal{V}^+&=&\frac{1}{2}\delta^+_{A,B}\mathcal{ZV}^+
 -\frac{1}{8}\Gamma^+_{ABC}\mathcal{ZV}^{+C}, \nonumber\\
\mathcal{D}^+_\mu\mathcal{D}^+_\nu\mathcal{V}^+&=&\frac{1}{2}\delta_{\mu\nu} 
 \mathcal{ZV}^+ -\frac{1}{2}\mathcal{ZV}^+_{\mu\nu} , \nonumber\\
\mathcal{D}^+ \mathcal{D}^+_\mu \mathcal{V}^+ &=& -\frac{i}{2}
 \partial_\mu  \mathcal{V}^+ -\frac{i}{2}
 \partial^\nu  \mathcal{V}^+_{\mu\nu},
 \nonumber\\
\mathcal{D}^+_\mu \mathcal{D}^+ \mathcal{V}^+ &=& -\frac{i}{2}
 \partial_\mu   \mathcal{V}^+  +\frac{i}{2}
 \partial^\nu   \mathcal{V}^+_{\mu\nu},\nonumber\\
\mathcal{D}^+_\mu \mathcal{D}^+_A \mathcal{V}^+ &=&  \frac{i}{2} \delta^+_{A,\mu\nu}
\partial^\nu  \mathcal{V}^+ +\frac{i}{2} \partial_\mu  
 \mathcal{V}^+_A -\frac{i}{8}\Gamma^+_{AB\mu\nu }
\partial^\nu  
 \mathcal{V}^{+B}, \nonumber\\
\mathcal{D}^+_A\mathcal{D}^+_\mu \mathcal{V}^+ &=&  \frac{i}{2}
 \delta^+_{A,\mu\nu} 
 \partial^\nu   \mathcal{V}^+ -\frac{i}{2}
 \partial_\mu  \mathcal{V}^+_A +\frac{i}{8}\Gamma^+_{AB\mu\nu
 }\partial^\nu  
 \mathcal{V}^{+B}. \label{eq:2nd differential relations of V}
\end{eqnarray}
We can also express eqs.(\ref{eq:2nd differential relations of V})
in terms of  $\mathcal{V}^+_A$ by using eqs.
(\ref{eq:constraints}). 
\begin{align}
\mathcal{V}^+| &=v , &\mathcal{V}^+_A| &= v^+_A , & \mathcal{ZV}^+ | &=
 K , & \mathcal{ZV}^+_A| &= K^+_A,\nonumber\\
\mathcal{D}^+_A \mathcal{V}^+| &= \lambda^+ _A ,&\mathcal{D}^+ \mathcal{V}^+| &=
 \lambda,& \mathcal{D}^+_\mu \mathcal{V}^+| &=\psi_\mu, && 
 \label{eq:lowest components of HP}
\end{align}
where $|$ means to take the lowest components of the $\theta$'s.
Higher components of $\theta$'s in the superfields $\mathcal{V}^+$ and
$\mathcal{V}^+_A$ are expressed by the derivative of the fields
(\ref{eq:lowest components of HP}). From
eqs. (\ref{eq:constraints}) and (\ref{eq:2nd differential relations of
V}), we show some twisted supertransformations:
\begin{eqnarray}
s^+ v^+_A &\equiv& \mathcal{Q}^+\mathcal{V}^+_A|  = \mathcal{D}^+\mathcal{V}^+_A|=-\lambda^+_A, \nonumber\\
s^+\lambda  &\equiv& \mathcal{Q}^+\mathcal{D}^+ \mathcal{V}^+|  =
 \mathcal{D}^+\mathcal{D}^+ \mathcal{V}^+| =\frac{1}{2}\mathcal{Z} \mathcal{V}^+|  = 
 \frac{1}{2}K.
\end{eqnarray}
We can find the other transformation laws of components fields. We
summarize the TSUSY transformations of the hypermultiplet in Table
\ref{tb:trhy1}. 

\begin{table}
\[
\begin{array}{|c||c|c|}
\hline
 & s^+ & s^+_\mu \\
\hline
v & \lambda & \psi^\mu\\
v^+_B & -\lambda ^+_B & -\delta^+_{B\mu\nu}\psi^\nu\\
\lambda &\frac{1}{2}K & -\frac{i}{2}\partial_\mu v
 +\frac{i}{2} \partial^\nu v^+_{\mu\nu}\\
\lambda^+_B & -\frac{1}{2}K^+_B & 
\frac{i}{2}\delta^+_{B,\mu\nu}\partial^\nu v
+\frac{i}{2}\partial_\mu v^+_B - \frac{i}{8}
\Gamma^+_{BC\mu\nu}\partial^\nu v^{+C}\\
\psi_\nu &  -\frac{i}{2}\partial_\nu v
 -\frac{i}{2} \partial^\rho v^+_{\nu\rho}&
 \frac{1}{2} \delta_{\mu\nu}K -\frac{1}{2}K^+_{\mu\nu}  \\
K & -i\partial^\mu \psi_\mu
& -i\partial_\mu \lambda +i\partial^\nu \lambda^+_{\mu\nu}\\
K^+_B&i\delta^+ _{B,\mu\nu}\partial^\mu \psi^\nu 
&i\delta^+_{B,\mu\nu}\partial^\nu\lambda+i\partial_\mu\lambda^+_B
+\frac{i}{4}\Gamma^+_{BC\mu\nu}\partial^\nu \lambda^{+C}\\
\hline
\end{array}
\]

\[
\begin{array}{|c||c|c|}
\hline
 & s^+_A & Z\\
\hline
v&\lambda^+_A & K \\
v^+_B &\delta^+_{A,B}\lambda+\frac{1}{4}\Gamma^+_{ABC}\lambda^{+C} &
 K^+_B\\ 
\lambda &\frac{1}{2}K^+_A&
 -i\partial^\mu\psi_\mu \\
\lambda^+_B &\frac{1}{2}\delta^+_{A,B}K
 -\frac{1}{8}\Gamma^+_{ABC}K^{+C}
& -i\delta^+_{B,\mu\nu}\partial^\mu\psi^\nu \\
\psi_\nu & \frac{i}{2}\delta^+_{A,\nu\rho}\partial^\rho
 v-\frac{i}{2}\partial_\nu v^+_A 
 +\frac{i}{8}\Gamma^+_{AB\nu\rho}\partial^\rho v^{+B} 
&-i\partial_\nu \lambda +i\partial^\rho\lambda^+_{\nu\rho} \\
K & -i\delta^+_{A,\mu\nu}\partial^\mu \psi^\nu
& -\partial^\mu \partial_\mu v\\
K^+_B &-i\delta^+_{A,B}\partial^\mu \psi_\mu -i
 \Gamma^+_{AB\mu\nu}\partial^\mu \psi^\nu & 
- \partial^\mu  \partial_\mu v^+_B\\
\hline
\end{array}
\]
\caption{$N$=2 twisted SUSY transformations of the Hypermultiplet.}
\label{tb:trhy1}
\end{table}

$N$=2 TSUSY with a central charge and R-invariant action is the
following form, 
\begin{eqnarray}
 \mathcal{S}_H
&=&\frac{1}{2} \int d^4x \mbox{Tr} \Big{(}
-\partial^\mu v  \partial_\mu v -\frac{1}{4}\partial^\mu
v^A \partial_\mu v_A  + 4i \psi^\mu
(\partial_\mu \lambda -\partial^\nu \lambda_{\mu\nu} ) \nonumber \\
 & &\hspace{25mm} + K^2+  \frac{1}{4} K^{+A} K^+ _A  \Big{)}.
\end{eqnarray}
This action can be represented by the superfields. We omit the explicit
form of the superfield action because of a complicated one. 
We will show a covariantized action with respect
to the superfields in the next section.

\section{Connection to the Vafa-Witten theory}
We pointed out that the hypermultiplet action coupling to the gauge
multiplet possess the twisted $N$=4 supersymmetry at on-shell level
\cite{KM}. We will investigate the $N$=4 twisted supersymmetry in this
case.  We define covariantized operators in order to introduce the gauge
multiplet.   
\begin{eqnarray}
\nabla^+&\equiv& \mathcal{D}^+ -i\Gamma, \nonumber\\
\nabla^+_\mu &\equiv& \mathcal{D}^+_\mu -i\Gamma_\mu, \nonumber\\
\nabla^+_A &\equiv& \mathcal{D}^+_A -i\Gamma_A, \nonumber\\
\nabla_{\underline{\mu}}&\equiv& \partial_\mu -i\Gamma_{\underline{\mu}},
\end{eqnarray}
where $\{\Gamma, \Gamma_\mu, \Gamma_A\}$ and $\Gamma_{\underline{\mu}}$
are fermionic and bosonic connection superfields, respectively. We then
use the superconnection formalism \cite{GSW,Sohnius,AL}. The
supercurvatures are defined as the following commutation relations,
\begin{align}
\{ \nabla^+,\nabla^+ _\mu\} &= -i\nabla_{\underline{\mu}} ,&
\{ \nabla^+ _A , \nabla^+ _\mu \} &= i\delta^+ _{A,\mu\nu}
 \nabla^{\underline{\nu}},& \{\nabla^+,\nabla^+_A\}&=0, \nonumber \\
\{\nabla^+,\nabla^+\} &= \mathcal{Z}-i\mathcal{F} ,&
 \{\nabla^+_\mu,\nabla^+_\nu\}& =\delta_{\mu\nu} (\mathcal{Z}-i\mathcal{W}) ,&
 \{\nabla^+_A,\nabla^+_B\} &=  \delta^+ _{A,B}
 (\mathcal{Z}-i\mathcal{F}), \nonumber\\
\{\nabla_{\underline{\mu}}, \nabla_{\underline{\nu}}\}&=
 -i\mathcal{F}_{\underline{\mu},\underline{\nu}}, & & & &
\end{align}
where $\mathcal{F}$ and $\mathcal{W}$ are bosonic curvature superfields
and appearing in the twisted vector multiplet and
$\mathcal{F}_{\underline{\mu},\underline{\nu}}$ are supercurvatures with
the gauge fields. The curvature
superfields $\mathcal{F}$, $\mathcal{W}$ and
$\mathcal{F}_{\underline{\mu},\underline{\nu}}$ commute with the central
charge $\mathcal{Z}$. 

In the covariant case the constraints of the superfield corresponding to
the constraints (\ref{eq:constraints}) are
\begin{eqnarray}
& &\delta^+_{A,\mu\nu}\nabla^{+\nu} \mathcal{V}^+
 +\nabla^+_\mu \mathcal{V}^+ _A =0, \nonumber\\
& &\nabla^+_A \mathcal{V}^+ +\nabla^+ \mathcal{V}^+_A =0,
 \nonumber\\
& &\nabla^+_A \mathcal{V}^+_B +\nabla^+_B \mathcal{V}^+_A
 -2\delta^+_{A,B} \nabla^+\mathcal{V}^+ =0, \nonumber\\
& &\nabla^+_A \mathcal{V}^+_B - \nabla^+_B \mathcal{V}^+_A
 -\frac{1}{4}\Gamma^+_{ABC}(\nabla^{+C}\mathcal{V}^+-\nabla^+
 \mathcal{V}^{+C} )=0. \label{eq:constraints non}
\end{eqnarray}
From the constraints (\ref{eq:constraints non}) we derive the following equations:
\begin{eqnarray}
\nabla^+\nabla^+\mathcal{V}^+&=&\frac{1}{2}\mathcal{ZV}^+
 +\frac{1}{2}[-i\mathcal{F}, \mathcal{V}^+], \nonumber\\
\nabla^+\nabla^+_A\mathcal{V}^+&=&-\frac{1}{2}\mathcal{ZV}^+_A
 -\frac{1}{2}[-i\mathcal{F},\mathcal{V}^+_A], \nonumber\\
\nabla^+_A\nabla^+\mathcal{V}^+&=&\frac{1}{2}\mathcal{ZV}^+_A
 +\frac{1}{2}[-i\mathcal{F},\mathcal{V}^+_A], \nonumber\\
\nabla^+_A\nabla^+_B\mathcal{V}^+&=&\frac{1}{2}\delta^+_{A,B}\mathcal{ZV}^+
 -\frac{1}{8}\Gamma^+_{ABC}\mathcal{ZV}^{+C}+\frac{1}{2}\delta^+_{A,B}
 [-i\mathcal{F},\mathcal{V}^+]-\frac{1}{8}\Gamma^+_{ABC}[-i\mathcal{F},
 \mathcal{V}^{+C}], \nonumber\\
\nabla^+_\mu\nabla^+_\nu\mathcal{V}^+&=&\frac{1}{2}\delta_{\mu\nu}
 \mathcal{ZV}^+ -\frac{1}{2}\mathcal{ZV}^+_{\mu\nu} +\frac{1}{2}
 \delta_{\mu\nu}[-i\mathcal{W},\mathcal{V}^+]-\frac{1}{2}[-i\mathcal{W},
 \mathcal{V}^+_{\mu\nu}], \nonumber\\
\nabla^+ \nabla^+_\mu \mathcal{V}^+ &=& -\frac{i}{2}
 \nabla_{\underline{\mu}}\mathcal{V}^+ -\frac{i}{2}
 \nabla^{\underline\nu}\mathcal{V}^+_{\mu\nu},
 \nonumber\\
\nabla^+_\mu \nabla^+ \mathcal{V}^+ &=& -\frac{i}{2}
 \nabla_{\underline{\mu}} \mathcal{V}^+  +\frac{i}{2}
 \nabla^{\underline\nu}\mathcal{V}^+_{\mu\nu},\nonumber\\
\nabla^+_\mu \nabla^+_A \mathcal{V}^+ &=&  \frac{i}{2} \delta^+_{A,\mu\nu}
 \nabla^{\underline\nu} \mathcal{V}^+ +\frac{i}{2} \nabla_{\underline\mu}
 \mathcal{V}^+_A -\frac{i}{8}\Gamma^+_{AB\mu\nu } \nabla^{\underline\nu}
 \mathcal{V}^{+B}, \nonumber\\
\nabla^+_A \nabla^+_\mu \mathcal{V}^+ &=&  \frac{i}{2} \delta^+_{A,\mu\nu}
 \nabla^{\underline\nu} \mathcal{V}^+ -\frac{i}{2} \nabla_{\underline\mu}
 \mathcal{V}^+_A +\frac{i}{8}\Gamma^+_{AB\mu\nu } \nabla^{\underline\nu}
 \mathcal{V}^{+B}. \label{eq:2nd differential relations of V non}
\end{eqnarray}
We define component fields of superfields which is consistent with
Abelian case.
\begin{align}
\mathcal{F}| &= \phi,  &\nabla^+_\mu \mathcal{F}| &= C_\mu, &
\frac{1}{4}\delta^+ _{A,\mu\nu} \nabla^{+\mu}\nabla^{+\nu}
\mathcal{F}| &=- \phi^+_{A},&& \nonumber \\
\mathcal{W}| &= \overline{\phi}, &\nabla^+_A \mathcal{W}| &=\chi^+ _A,
&\nabla^+ \mathcal{W}| &=\chi, &&\nonumber \\
\mathcal{V}^+| &=v , &\mathcal{V}^+_A| &= v^+_A , & \mathcal{ZV}^+ | &=
 K , & \mathcal{ZV}^+_A| &= K^+_A, \nonumber\\
\nabla^+_A \mathcal{V}^+| &= \lambda^+ _A ,& \nabla^+ \mathcal{V}^+| &=
 \lambda,& \nabla^+_\mu \mathcal{V}^+| &=\psi_\mu, && \nonumber\\
\nabla_{\underline{\mu}}| &= \mathcal{D}_{\mu}, &
 \Gamma_{\underline{\mu}}| &= A_\mu, & & & &
 \label{eq:lowest components}
\end{align}
where $\mathcal{D}_\mu$ and $A_\mu$ stand for the usual covariant
derivatives and the gauge fields, respectively.
Taking into account the Wess-Zumino gauge, we eliminate the
lowest components of the fermionic superconnections and some fields
appearing in higher components of $\theta's$. For example a nontrivial
TSUSY transformation can be derived in the following manner, 
\begin{eqnarray}
s^+\lambda &\equiv&  \mathcal{Q}^+\nabla^+
 \mathcal{V}^+|=\nabla^+\nabla^+ \mathcal{V}^+|  \nonumber \\
&=&\frac{1}{2}\mathcal{Z}\mathcal{V}^+|-\frac{i}{2}\{\mathcal{F}, \mathcal{V}^+\}| = \frac{1}{2}K-\frac{i}{2}[\phi, v].
\end{eqnarray}
We can also find the other twisted supertransformations for all the
component fields. We show the $N=2$ twisted SUSY
transformations of the hypermultiplet in Table \ref{tb:trhy1 non}.

\begin{table}
\[
\begin{array}{|c||c|c|}
\hline
 & s^+ & s^+_\mu \\
\hline
v & \lambda & \psi^\mu\\
v^+_B & -\lambda ^+_B & -\delta^+_{B\mu\nu}\psi^\nu\\
\lambda &\frac{1}{2}K -\frac{i}{2}[\phi,v]& -\frac{i}{2}\mathcal{D}_\mu v
 +\frac{i}{2} \mathcal{D}^\nu v^+_{\mu\nu}\\
\lambda^+_B & -\frac{1}{2}K^+_B + \frac{i}{2}[\phi,v^+_A]& 
\frac{i}{2}\delta^+_{B,\mu\nu}\mathcal{D}^\nu v
+\frac{i}{2}\mathcal{D}_\mu v^+_B - \frac{i}{8}
\Gamma^+_{BC\mu\nu}\mathcal{D}^\nu v^{+C}\\
\psi_\nu &  -\frac{i}{2}\mathcal{D}_\nu v
 -\frac{i}{2} \mathcal{D}^\rho v^+_{\nu\rho}&
 \frac{1}{2} \delta_{\mu\nu}K -\frac{1}{2}K^+_{\mu\nu} 
-\frac{i}{2}\delta_{\mu\nu} [\overline{\phi},v]
+\frac{i}{2}[\overline{\phi},v^+_{\mu\nu}] \\
K & -i\mathcal{D}^\mu \psi_\mu
 +\frac{i}{2}[\chi,v]
& -i\mathcal{D}_\mu \lambda +i\mathcal{D}^\nu \lambda^+_{\mu\nu}
+\frac{i}{2} [C_\mu,v]\\
 &+\frac{i}{8}[\chi^+_{A},v^{+A}] +i[\overline{\phi},\lambda]   
&  +\frac{i}{2} [C^\nu,v^+_{\mu\nu}
]+i[\phi,\psi_\mu] \\
K^+_B&i\delta^+ _{B,\mu\nu}\mathcal{D}^\mu \psi^\nu 
- \frac{i}{2}[\chi^+_B,v]+\frac{i}{2}[\chi,v^+_B]
&i\delta^+_{B,\mu\nu}\mathcal{D}^\nu\lambda+i\mathcal{D}_\mu\lambda^+_B
+\frac{i}{4}\Gamma^+_{BC\mu\nu}\mathcal{D}^\nu \lambda^{+C}       \\

& -\frac{i}{32}\Gamma^+_{BCD}[\chi^{+C},v^{+D}] -i[\overline{\phi},\lambda^+_B]&
 -\frac{i}{2}\delta^+_{B,\mu\nu}[C^\nu,v] +\frac{i}{2}[C_\mu,v^+_B] \\
& &+\frac{i}{8}\Gamma^+_{BC\mu\nu}[C^\nu,v^{+C}]-i\delta^+_{B,\mu\nu}[\phi,\psi^\nu] \\
\hline
\end{array}
\]
\[
\begin{array}{|c||c|c|}
\hline
 & s^+_A \\
\hline
v&\lambda^+_A  \\
v^+_B &\delta^+_{A,B}\lambda+\frac{1}{4}\Gamma^+_{ABC}\lambda^{+C} \\ 
\lambda &\frac{1}{2}K^+_A-\frac{i}{2}[\phi,v^+_A]\\
\lambda^+_B &\frac{1}{2}\delta^+_{A,B}K
 -\frac{1}{8}\Gamma^+_{ABC}K^{+C}-\frac{i}{2}\delta^+_{A,B}[\phi,v]
+\frac{i}{8}\Gamma^+_{ABC}[\phi,v^{+C}]      \\
\psi_\nu & +\frac{i}{2}\delta^+_{A,\nu\rho}\mathcal{D}^\rho
 v-\frac{i}{2}\mathcal{D}_\nu v^+_A 
 +\frac{i}{8}\Gamma^+_{AB\nu\rho}\mathcal{D}^\rho v^{+B}  \\
K & -i\delta^+_{A,\mu\nu}\mathcal{D}^\mu \psi^\nu
 -\frac{i}{2}[\chi,v^+_A]
+\frac{i}{2}[\chi^+_A,v]+\frac{i}{32}\Gamma^+_{ABC}[\chi^{+B},v^{+C}]+i[\overline{\phi},\lambda^+_A]\\
K^+_B &-i\delta^+_{A,B}\mathcal{D}^\mu \psi_\mu -i
 \Gamma^+_{AB\mu\nu}\mathcal{D}^\mu \psi^\nu
 +\frac{i}{2}[\chi^+_A,v^+_B] -\frac{i}{2}[\chi^+_B,v^+_A]
 +\frac{i}{2}\delta^+_{A,B}[\chi,v] \\
& +\frac{i}{8}\delta^+_{A,B}[\chi^{+C},v^+_C]
 +\frac{i}{8}\Gamma^+_{ABC}[\chi^{+C},v]
 -\frac{i}{8}\Gamma^+_{ABC}[\chi,v^{+C}]   \\
&+i\delta^+_{A,B}[\overline{\phi},\lambda]+\frac{i}{4}\Gamma^+_{ABC}[\overline{\phi},\lambda^{+C}] \\
\hline
\end{array}
\]
\[
\begin{array}{|c||c|}
\hline
 & Z\\
\hline
v& K \\
v^+_B  &
 K^+_B\\ 
\lambda&
 -i\mathcal{D}^\mu\psi_\mu
 +\frac{i}{2}[\chi,v]+\frac{i}{8}[\chi^{+A},v^+_A]+i[\overline{\phi},\lambda]\\
\lambda^+_B 
& -i\delta^+_{B,\mu\nu}\mathcal{D}^\mu\psi^\nu +\frac{i}{2}[\chi^+_B,v]
-\frac{i}{2}[\chi,v^+_B] +\frac{i}{32}\Gamma^+_{BCD}[\chi^{+C},v^{+D}]+i[\overline{\phi},\lambda^+_B] \\
\psi_\nu 
&-i\mathcal{D}_\nu \lambda +i\mathcal{D}^\rho\lambda^+_{\nu\rho}
+\frac{i}{2}[C_\nu,v]+\frac{i}{2}[C^\rho,v^+_{\nu\rho}]+i[\phi,\psi_\nu]     \\
K & -\mathcal{D}^\mu \mathcal{D}_\mu v
+i\{C^\mu,\psi_\mu\}+\frac{i}{4}\{\chi^{+A},\lambda^+_A\}
+i\{\chi,\lambda\}-\frac{i}{4}[\phi^{+A},v^+_A ] \\
& +i[\phi,K]+i[\overline{\phi},K]
 +\frac{1}{2}[\phi,[\overline{\phi},v]]+\frac{1}{2}[\overline{\phi},[\phi,v]]\\  
K^+_B  & -
 \mathcal{D}^\mu  \mathcal{D}_\mu v^+_B
 -i\delta^+_{B,\mu\nu}\{C^\mu,\psi^\nu\} +i\{\chi^+_B,\lambda \}
-i\{\chi,\lambda^+_B \}\\
& -\frac{i}{16}
 \Gamma^+_{BCD}\{\chi^{+C},\lambda^{+D}\}+i[\phi^+_B,v] +i[\phi,K^+_B] +i[\overline{\phi},K^+_B]\\
 &
 +\frac{i}{16}\Gamma^+_{BCD}[\phi^{+C},v^{+D}]+\frac{1}{2}[\phi,[\overline{\phi},v^+_B]] 
 + \frac{1}{2}[\overline{\phi},[\phi,v^+_B]]\\
\hline
\end{array}
\]
\caption{$N$=2 twisted SUSY transformations of the Hypermultiplet
 coupling to the vector multiplet.}
\label{tb:trhy1 non}
\end{table}

These transformations satisfy the following commutation relations,
\begin{align}
\{s^+,s^+\}\varphi &= Z\varphi -i[\phi,\varphi],& \{s^+,s^+_\mu\}\varphi
 &= -iD_\mu \varphi,& {[}s^+,Z{]} \varphi &= 0, \nonumber\\
\{s^+_\mu,s^+_\nu\}\varphi &=\delta_{\mu\nu} Z\varphi -i\delta_{\mu\nu}
 [\overline{\phi},\varphi],& \{s^+_\mu,s^+_A\}\varphi &=
 i\delta^+ _{\mu\nu,A} D^\nu\varphi,&
 {[}s^+_\mu,Z{]} \varphi &= 0, \nonumber \\
\{s^+_A,s^+_B\} \varphi &= \delta^+ _{A,B} Z \varphi -i\delta^+
 _{A,B}[\phi,\varphi],& \{s^+,s^+_A\}\varphi &= 0,
& {[}s^+_A,Z{]} \varphi &= 0,
\end{align}
where $\varphi = v, v^+_A, \lambda , \lambda^+_A, \psi_\mu ,
K, K^+_A$ and $D_{\mu}\varphi =\partial_\mu \varphi -i[\omega_\mu,
\varphi]$. These algebras are closed at the off-shell level up to the
gauge transformation.

We can construct a covariantized action of the hypermultiplet. We derive 
the action by taking the lowest components of the $\theta$'s. We then
obtain the covariantized action as follows:
{
\begin{eqnarray}
 \mathcal{S}_H
&=& \int d^4x \mbox{Tr}\nonumber \\
&  \times &\hspace{0mm}\Biggl{\{}  -\frac{1}{12} \Big{(} \frac{1}{16}\Gamma^+_{AB\mu\nu}
 \nabla^{+\mu} \nabla^{+\nu} (\mathcal{V}^{+A}\mathcal{ZV}^{+B})+ \nabla^{+\mu} \nabla^{+\nu}
 (\mathcal{V}^+\mathcal{ZV}^+_{\mu\nu}- 
 \mathcal{V}^+_{\mu\nu}\mathcal{ZV}^+) \Big{)} \nonumber\\
& &\hspace{0mm}+ \frac{1}{12} \Big{(} \frac{1}{64} \Gamma^+_{ABC}
 \nabla^{+A} 
 \nabla^{+B} (\mathcal{V}^{+C}\mathcal{ZV}^+)
- \frac{1}{64}
 \Gamma^+_{ABC}\nabla^{+A} \nabla^{+B} (\mathcal{V}^+\mathcal{ZV}^{+C})
 \nonumber\\
& &\hspace{0mm} + \frac{1}{64} \Gamma^+_{ABC} \nabla^+ \nabla^{+A}
 (\mathcal{V}^{+B}\mathcal{ZV}^{+C})
- \frac{1}{64} \Gamma^+_{ABC} \nabla^{+A} \nabla^+
(\mathcal{V}^{+B}\mathcal{ZV}^{+C})\Big{)} \nonumber\\
& &\hspace{0mm}+\frac{1}{12}\Big{(} \frac{1}{4} \nabla^{+A}\nabla^+
 (\mathcal{V}^+\mathcal{ZV}^+_A)+ \frac{1}{4} \nabla^+
\nabla^{+A}(\mathcal{V}^+_A\mathcal{ZV}^+)+\frac{1}{16}\nabla^{+A}
\nabla^{+B} (\mathcal{V}^+_B\mathcal{ZV}^+_A) \nonumber\\
& &\hspace{0mm}- \frac{1}{4}
 \nabla^{+A}\nabla^+ (\mathcal{V}^+_A\mathcal{ZV}^+)- \frac{1}{4} \nabla^+
\nabla^{+A}(\mathcal{V}^+\mathcal{ZV}^+_A)-\frac{1}{16}\nabla^{+A}\nabla^{+B}
(\mathcal{V}^+_A\mathcal{ZV}^+_B)\Big{)} \Biggl{\}}\Biggl{|}
 \nonumber\\
&=&\frac{1}{2} \int d^4x \mbox{Tr} \nonumber \\
&\times  &\hspace{0mm} \Big{(}-\mathcal{D}^\mu v  \mathcal{D}_\mu v -\frac{1}{4}\mathcal{D}^\mu
v^A\mathcal{D}_\mu v_A +K^2+  \frac{1}{4}K^{+A}K^+_{A}+4i \psi^\mu
(\mathcal{D}_\mu \lambda -\mathcal{D}^\nu \lambda_{\mu\nu} )\nonumber\\
& &\hspace{0mm}
+ 2i\phi\{\psi^\mu,\psi_\mu\}+2i\overline{\phi}\{\lambda,\lambda\}
+\frac{i}{2}\overline{\phi}\{\lambda^{+A},\lambda^+_A\}
-2iv\{C^\mu,\psi_\mu\} -2iv\{\chi,\lambda\} \nonumber\\
& &\hspace{0mm}
-\frac{i}{2}v\{\chi^{+A},\lambda^+_A\}
-\frac{i}{2}v^{+A}\{\chi^+_A,\lambda\}
+\frac{i}{2}v^{+A}\{\chi,\lambda^+_A\}
+2iv^+_{\mu\nu}\{C^\mu,\psi^\nu\}    \nonumber\\
& &\hspace{0mm}
+\frac{i}{32}\Gamma^{+ABC}v^+_A \{\chi^+_B,\lambda^+_C\}
-\frac{i}{64}\Gamma^{+ABC}\phi^+_A
[v^+_B,v^+_C]+\frac{i}{2}\phi^+_A[v^+_A,v] \nonumber\\
& &\hspace{0mm}
-\frac{1}{2}v[\phi,[\overline{\phi},v]]
-\frac{1}{2}v[\overline{\phi},[\phi,v]]
-\frac{1}{8}v^{+A}[\phi,[\overline{\phi},v^+_A]] 
-\frac{1}{8}v^{+A}[\overline{\phi},[\phi,v^+_A]] \Big{)}.
\end{eqnarray}}
The action is invariant for all the twisted supertransformations,
but the action cannot be expressed by $\{s^+, s^+_\mu, s^+_A\}$ exact
form.  
We construct an off-shell Donaldson-Witten theory coupled to the twisted
hypermultiplet. The off-shell Donaldson-Witten theory are given 
by using the twisted superspace formalism \cite{AL,KKM}:
\begin{eqnarray}
S_{DW} &=& \frac{1}{2} \int d^4x d^4\theta\ \mbox{Tr} \mathcal{F}^2
 \nonumber \\
&=&\frac{1}{2} \int d^4 x\text{Tr}\Big{(}-\phi D^\mu D_\mu
 \overline{\phi} -iC^\mu(D_\mu \chi -D^\nu \chi^+ _{\mu\nu} )  + (F^-
 _{\mu\nu})^2 \nonumber \\
& &\hspace{23mm}+\frac{i}{2}\phi \{\chi,\chi\}+\frac{i}{8}\phi
 \{\chi^{+A},\chi^+ _A\} + \frac{i}{2}\overline{\phi}
 \{C^\mu,C_\mu\}\nonumber  \\
& &\hspace{23mm} +\frac{1}{4}[\phi,\overline{\phi}]^2
 -\frac{1}{4}(\phi^+ _{\mu\nu} )^2 \Big{)}.
\end{eqnarray}
We redefine the component fields as
\begin{eqnarray}
\lambda\rightarrow \frac{1}{2}\lambda,\qquad \lambda^+_A \rightarrow
\frac{1}{2}\lambda^+_A ,\qquad \psi_\mu \rightarrow
-\frac{1}{2}\psi_\mu,\qquad \overline{\phi}\rightarrow
-\overline{\phi},
\end{eqnarray}
in order to adjust coefficients in the action. We then derive the
off-shell Donaldson-Witten theory coupled to the twisted hypermultiplet
as follows:
\begin{eqnarray}
\mathcal{S}&=&-(\mathcal{S}_H+\mathcal{S}_{DW} )\nonumber\\
&=&-\frac{1}{2} \int d^4x \mbox{Tr} \Big{(}
-\mathcal{D}^\mu v  \mathcal{D}_\mu v
-\frac{1}{4}\mathcal{D}^\mu v^A\mathcal{D}_\mu v_A
+\phi D^\mu D_\mu \overline{\phi} + (F^- _{\mu\nu})^2 \nonumber\\
& &\hspace{20mm}
-i \psi^\mu (\mathcal{D}_\mu \lambda -\mathcal{D}^\nu \lambda_{\mu\nu} )
+ \frac{i}{2}\phi\{\psi^\mu,\psi_\mu\}
-\frac{i}{2}\overline{\phi}\{\lambda,\lambda\}
-\frac{i}{8}\overline{\phi}\{\lambda^{+A},\lambda^+_A\}
 \nonumber\\
& &\hspace{20mm}
-iC^\mu(D_\mu \chi -D^\nu \chi^+ _{\mu\nu} )
- \frac{i}{2} \overline{\phi} \{C^\mu,C_\mu\}
+\frac{i}{2}\phi \{\chi,\chi\}
+\frac{i}{8}\phi \{\chi^{+A},\chi^+ _A\}
\nonumber\\
& &\hspace{20mm}
-\frac{i}{4}v\{\chi^{+A},\lambda^+_A\}
+\frac{i}{64}\Gamma^{+ABC}v^+_A \{\chi^+_B,\lambda^+_C\}
-iv^+_{\mu\nu}\{C^\mu,\psi^\nu\}
 \nonumber\\
& &\hspace{20mm}
 +iv\{C^\mu,\psi_\mu\}  
-\frac{i}{4}v^{+A}\{\chi^+_A,\lambda\}
+\frac{i}{4}v^{+A}\{\chi,\lambda^+_A\}
 -iv\{\chi,\lambda\}  \nonumber\\
& &\hspace{20mm}
 +\frac{1}{2}v[\phi,[\overline{\phi},v]]
+\frac{1}{2}v[\overline{\phi},[\phi,v]]
+\frac{1}{8}v^{+A}[\phi,[\overline{\phi},v^+_A]]
 \nonumber\\
& &\hspace{20mm}
+\frac{1}{8}v^{+A}[\overline{\phi},[\phi,v^+_A]]
 +\frac{1}{4}[\phi,\overline{\phi}]^2
 +\frac{i}{2}\phi^{+A}[v^+_A, v] \nonumber  \\
 & &\hspace{20mm}
-\frac{i}{64}\Gamma^{+ABC}\phi^+_A[v^+_B,v^+_C]
+KK+\frac{1}{4}K^{+A}K^+_A -\frac{1}{4}\phi^{+A}
 \phi^+_A
\Big{)}. \label{eq: off-shell DW coupled to hypermultiplet}
\end{eqnarray} 
This action possesses the off-shell $N$=2 twisted supersymmetry.
We then investigate a symmetry when we eliminate the auxiliary fields $K, K^+_A$
and $\phi^+_A$. The equations of motion of the auxiliary fields are
\begin{eqnarray}
\phi^+_A = -\frac{i}{32}\Gamma^+_{ABC}[v^{+B},v^{+C}]+i[v^+_A,v],
\qquad K=K^+_A=0.
\end{eqnarray}
We find that the $Z$ transformations for the components $\{v, v^+_A,
\lambda, \lambda^+_A, \psi_\mu\}$ of the twisted hypermultiplet are
equivalent to the on-shell conditions and therefore these
transformations disappear at the on-shell level. The on-shell
action is given by the following form:
\begin{eqnarray}
\mathcal{S}_{\mbox{on-shell}}
&=&-\frac{1}{2} \int d^4x \mbox{Tr}\nonumber \\
& \times &\Big{(}
-\mathcal{D}^\mu v  \mathcal{D}_\mu v
-\frac{1}{4}\mathcal{D}^\mu v^A\mathcal{D}_\mu v_A
+\phi D^\mu D_\mu \overline{\phi} + (F^- _{\mu\nu})^2 \nonumber\\
& &
-i \psi^\mu (\mathcal{D}_\mu \lambda -\mathcal{D}^\nu \lambda_{\mu\nu} )
+ \frac{i}{2}\phi\{\psi^\mu,\psi_\mu\}
-\frac{i}{2}\overline{\phi}\{\lambda,\lambda\}
-\frac{i}{8}\overline{\phi}\{\lambda^{+A},\lambda^+_A\}
 \nonumber\\
& &
-iC^\mu(D_\mu \chi -D^\nu \chi^+ _{\mu\nu} )
- \frac{i}{2} \overline{\phi} \{C^\mu,C_\mu\}
+\frac{i}{2}\phi \{\chi,\chi\}
+\frac{i}{8}\phi \{\chi^{+A},\chi^+ _A\}
\nonumber\\
& &
-\frac{i}{4}v\{\chi^{+A},\lambda^+_A\}
+\frac{i}{64}\Gamma^{+ABC}v^+_A \{\chi^+_B,\lambda^+_C\}
-iv^+_{\mu\nu}\{C^\mu,\psi^\nu\}
   \nonumber\\
& &
 +iv\{C^\mu,\psi_\mu\}
-\frac{i}{4}v^{+A}\{\chi^+_A,\lambda\}
+\frac{i}{4}v^{+A}\{\chi,\lambda^+_A\}
 -iv\{\chi,\lambda\}  \nonumber\\
& &
+\frac{1}{2}v[\phi,[\overline{\phi},v]]
+\frac{1}{2}v[\overline{\phi},[\phi,v]]
+\frac{1}{8}v^{+A}[\phi,[\overline{\phi},v^+_A]]
+\frac{1}{8}v^{+A}[\overline{\phi},[\phi,v^+_A]] \nonumber\\
& &
-\frac{1}{32}[v^+_A,v^+_B][v^{+A},v^{+B}]-\frac{1}{4}[v^+_A,v][v^{+A},v]
\Big{)}. \label{eq: on-shell DW coupled to hypermultiplet}
\end{eqnarray}
We then find that the action (\ref{eq: on-shell DW coupled to
hypermultiplet}) possesses the following symmetries:
\begin{align}
\phi &\rightarrow -\overline{\phi}, &  \overline{\phi} &\rightarrow
 -\phi, & v  &\rightarrow  v, & v^+_A &\rightarrow -v^+_A, \nonumber\\
 \chi  &\leftrightarrow \lambda, & \chi^+_A &\leftrightarrow
 \lambda^+_A, & C_\mu &\leftrightarrow \lambda_\mu , & A_\mu
 &\rightarrow A_\mu. 
\end{align}
Applying these discrete symmetries to the twisted SUSY transformations,
we obtain new fermionic symmetries which is shown in \ref{ch:N=4
symmetry}. We express new scalar, vector and anti-self-dual tensor
supercharges as $\{\overline{s}^+, \overline{s}^+_\mu,
\overline{s}^+_A\}$. We then find that the commutation relations
of the supercharge form a new $N$=4 twisted SUSY algebra and this
action possesses the $N=4$ twisted supersymmetry.  
The explicite form of the $N$=4 twisted SUSY algebras are given in
\ref{ch:N=4 symmetry}. 
These algebras closed at the on-shell level up to the gauge
transformations. In this case the ghost numbers of twisted scalar 
supercharges $s^+$ and $\overline{s}^+$  are $+1$ and $-1$, respectively. 
$s^+$ is BRST charge and $\overline{s}^+$ is anti-BRST charge.
We conclude that this $N$=4 twisted supersymmetric theory is equivalent
to the Vafa-Witten theory.

\section{$N=4$ twisted superconnection formalism}
The Marcus type $N$=4 twisted supersymmetric theory was derived from
 $N$=4 twisted superconnection formalism \cite{N-D}. 
We then reconstruct the $N=4$ twisted super Yang-Mills theory (\ref{eq:
 on-shell DW coupled to hypermultiplet}) which is derived from the $N$=2
 superspace formalism by using $N$=4 twisted superconnection formalism,
similarly. 
We represent sixteen supercovariant derivatives 
as $\{\nabla^+,
\nabla^+_\mu, \nabla^+_A, \overline{\nabla}^+, \overline{\nabla}^+_\mu,
\overline{\nabla}^+_A\}$.
We impose special constraints on general curvature superfields based on
the twisted superalgebras in \ref{ch:N=4 symmetry}, and
construct the $N=4$ twisted super Yang-Mills theory directly. 
We define the (anti)commutation relations of these
supercovariant derivatives in Table \ref{tb:N=4 supercurvatures}, where
$\mathcal{F}$, $\mathcal{W}$, $\mathcal{V}^+$ and $\mathcal{V}^+_A$ are
bosonic supercurvatures,
$\mathcal{F}_{\underline{\mu}}$, $\mathcal{F}_{A,\underline{\nu}}$,
$\overline{\mathcal{F}}_{\mu,\underline{\nu}}$,
$\overline{\mathcal{F}}_{\underline{\nu}}$,
$\overline{\mathcal{F}}_{A,\underline{\nu}}$ and 
$\mathcal{F}_{\mu,\underline{\nu}}$ are fermionic supercurvatures and
$\mathcal{F}_{\underline{\mu}, \underline{\nu}}$ is a bosonic
supercurvature which includes the ordinary curvature
$F_{\mu\nu}$. We can gauge away some superfluous fields in the
superconnections by taking Wess-Zumino gauge.
\begin{table}
\[\hspace{0mm}
{\footnotesize
\begin{array}{|c|ccc|ccc|c|}
\hline
 & \nabla^+ & \nabla^+_B &\overline{\nabla}^+_\nu &\overline{\nabla}^+
&\overline{\nabla}^+_B & \nabla^+_\nu & \nabla_{\underline{\nu}}\\
\hline
\nabla^+ & -i\mathcal{F} & 0 & 0 &-i\mathcal{V}^+ & -i\mathcal{V}^+_B &
-i\nabla_{\underline{\nu}} &   -i\mathcal{F}_{\underline{\nu}}\\
\nabla^+_A && -i\delta^+_{A,B}\mathcal{F} & 0 & i\mathcal{V}^+_A&
-i\delta^+_{A,B}\mathcal{V}^+
-\frac{i}{4}\Gamma^+_{ABC}\mathcal{V}^{+C} &
i\delta^+_{A,\nu\rho}\nabla^{\underline{\rho}} &
-i\mathcal{F}_{A,\underline{\nu}}\\
\overline{\nabla}^+_\mu & & & -i\delta_{\mu\nu} \mathcal{F}
 &-i\nabla_{\underline{\mu}} &
 i\delta^+_{B,\mu\nu}\nabla^{\underline{\nu}}
 &i\delta_{\mu\nu}\mathcal{V}^+-i\mathcal{V}^+_{\mu\nu} &
 -i\overline{F}_{\mu\underline{\nu}}\\
\hline
\overline{\nabla}^+ & & & &-i\mathcal{W} &0 & 0
&-i\overline{\mathcal{F}}_{\underline{\nu}} \\
\overline{\nabla}_A & & & & & -i\delta^+_{A,B}\mathcal{W} &  0 &
 -i\overline{F}_{A,\underline{\nu}} \\
\nabla^+_\mu & & & & & &-i\delta_{\mu\nu}\mathcal{W}
 &-i\mathcal{F}_{\mu\underline{\nu}} \\
\hline
\nabla_{\underline{\mu}} & & & & & & &
-i\mathcal{F}_{\underline{\mu}\underline{\nu}}\\
\hline
\end{array}}
\]
\caption{$N$=4 twisted supercurvatures}
\label{tb:N=4 supercurvatures}
\end{table}
From Jacobi identities of the supercovariant derivatives, we derive the
following nontrivial relations:
\begin{eqnarray}
\nabla^+ \mathcal{F}&=& \nabla^+_A \mathcal{F}= \overline{\nabla}^+_\mu
 \mathcal{F} = 0, \nonumber\\
\overline{\nabla}^+ \mathcal{W}&=& \overline{\nabla}^+_A \mathcal{W}
= \nabla^+_\mu \mathcal{W} = 0, \label{(anti)chiral-like conditions}
\end{eqnarray}
\begin{align}
\mathcal{F}_{\underline{\mu}} &= -\frac{i}{2}\nabla^+_\mu \mathcal{F},&
 \mathcal{F}_{A,\underline{\mu}} &=
-\frac{i}{2}\delta^+_{A,\mu\nu} \nabla^{+\nu} \mathcal{F},&
 \mathcal{F}_{\mu \underline{\nu}}
 &= -\frac{i}{2}\delta_{\mu\nu} \overline{\nabla}^+ \mathcal{F}
+\frac{i}{2} \overline{\nabla}^+_{\mu\nu} \mathcal{F},
 \nonumber\\
 \overline{ \mathcal{F}}_{\underline{\mu}} &= -\frac{i}{2}
 \overline{\nabla}^+_\mu \mathcal{W},
&\overline{ \mathcal{F}}_{A,\underline{\mu}}
 &= -\frac{i}{2}\delta^+_{A,\mu\nu}  \overline{\nabla}^{+\nu} \mathcal{W},
&\overline{ \mathcal{F}}_{\mu \underline{\nu}}
 &= -\frac{i}{2}\delta_{\mu\nu}\nabla^+ \mathcal{W}
 +\frac{i}{2}\nabla^+_{\mu\nu} \mathcal{W}, \label{fermionic supercurvature}
\end{align}
\begin{eqnarray}
\mathcal{F}_{\underline{\mu}\underline{\nu}} &=&
 -\frac{1}{2}\nabla^+\nabla^+_{\mu\nu} \mathcal{W}
+\frac{1}{4}[\nabla^+_\mu,\nabla^+_\nu] \mathcal{F} \nonumber\\
&=&
 -\frac{1}{2}\overline{\nabla}^+\overline{\nabla}^+_{\mu\nu} \mathcal{F}
+\frac{1}{4}[\overline{\nabla}^+_\mu,\overline{\nabla}^+_\nu]
\mathcal{W}, \label{bosonic supercurvature}
\end{eqnarray}
\begin{align}
 \nabla^+\mathcal{V}^+ &=-\frac{1}{2}\overline{\nabla}^+\mathcal{F},
 &\overline{\nabla}^+_{\mu}\mathcal{V}^+ &= \frac{1}{2}\nabla^+_\mu
 \mathcal{F},& \nabla^+_A\mathcal{V}^+
 &=-\frac{1}{2}\overline{\nabla}^+_A\mathcal{F}, \nonumber\\
 \nabla^+ \mathcal{V}^+_A &= -\frac{1}{2}\overline{\nabla}^+_A \mathcal{F},
&\nabla^+_\mu \mathcal{V}^+_A &= -\frac{1}{2}\delta^+_{A,\mu\nu} \nabla^{+\nu} \mathcal{F},
 \nonumber\\
 \overline{\nabla}^+\mathcal{V}^+ &= -\frac{1}{2}\nabla^+\mathcal{W},
 &\nabla^+_\mu\mathcal{V}^+ &= \frac{1}{2}\overline{\nabla}^+_\mu
 \mathcal{W},&\overline{\nabla}^+_A\mathcal{V}^+&=
 -\frac{1}{2}\nabla^+_A\mathcal{W}, \nonumber\\
 \overline{\nabla}^+ \mathcal{V}^+_A &= \frac{1}{2}\nabla^+_A \mathcal{W},
 & \overline{\nabla}^+_\mu \mathcal{V}^+_A &=
 \frac{1}{2}\delta^+_{A,\mu\nu}\overline{\nabla}^{+\nu} \mathcal{W},
\nonumber 
\end{align}
\vspace{-8mm}
\begin{eqnarray}
 \nabla^+_A \mathcal{V}^+_B &=& \frac{1}{2}\delta^+_{A,B}
 \overline{\nabla}^+ \mathcal{F}+\frac{1}{8}\Gamma^+_{ABC}
 \overline{\nabla}^{+C}\mathcal{F},\nonumber \\
 \overline{\nabla}^+_A \mathcal{V}^+_B &=&- \frac{1}{2}\delta^+_{A,B}
 \nabla^+ \mathcal{W}-\frac{1}{8}\Gamma^+_{ABC} \nabla^{+C}
 \mathcal{W},\nonumber
\label{higher components of V and V_A}
\end{eqnarray}
It should be noted that the curvature superfields $\mathcal{F}$ and
$\mathcal{W}$ are (anti)chiral superfield of non-Abelian type from
eq. (\ref{(anti)chiral-like conditions}).  
$\mathcal{F}_{\underline{\mu}}$, $\mathcal{F}_{A,\underline{\mu}}$,
$\mathcal{F}_{\mu,\underline{\nu}}$,
$\overline{\mathcal{F}}_{\underline{\mu}}$,
$\overline{\mathcal{F}}_{A,\underline{\mu}}$ and 
$\overline{\mathcal{F}}_{\mu,\underline{\nu}}$ are expressed by the
superfields $\mathcal{F}$ and $\mathcal{W}$.
The following useful equations
are then derived from the Jacobi identities
\begin{eqnarray}
\overline{\nabla}^+ \nabla^+_\mu \mathcal{F} &=&-i\nabla_{\underline{\mu}}
 \mathcal{V}^+ -i\nabla^{\underline{\nu}} \mathcal{V}^+_{\mu\nu},
 \nonumber\\
 \nabla^+_\mu \overline{\nabla}^+_A \mathcal{F} &=&-i
 \delta^+_{A,\mu\nu}\nabla^{\underline{\nu}}\mathcal{V}^+
 +i \nabla_{\underline{\mu}} \mathcal{V}^+_A -\frac{i}{4}
 \Gamma^+_{AB\mu\nu}\overline{\nabla}^{+\nu}\mathcal{V}^{+B}, \nonumber\\
 \overline{\nabla}^+_\mu \nabla^+_\nu \mathcal{F} &=&
 i\delta_{\mu\nu}[\mathcal{V}^+,\mathcal{F}]
 -i[\mathcal{V}^+_{\mu\nu},\mathcal{F}], \nonumber\\
 \nabla^+_A \overline{\nabla}^+_B \mathcal{F}&=& -i\delta^+_{A,B}
 [\mathcal{V}^+, \mathcal{F}]-\frac{i}{4}\Gamma^+_{ABC}
 [\mathcal{V}^{+C},\mathcal{F}], \nonumber\\
\overline{\nabla}^+\overline{\nabla}^+_A \mathcal{F}&=&
 \frac{i}{32}\Gamma^+_{ABC}[\mathcal{V}^{+B},  \mathcal{V}^{+C}]-i
 [ \mathcal{V}^+_A , \mathcal{V}^+]-2 \mathcal{F}^+_{\underline{A}},
 \nonumber\\
\overline{\nabla}^+_A\overline{\nabla}^+_B
 \mathcal{F}&=&-\frac{i}{4}\Gamma^+_{ABC}[\mathcal{V}^{+C},\mathcal{V}^+
 ]+i[\mathcal{V}^+_A,\mathcal{V}^+_B ] +\frac{1}{2}\Gamma^+_{ABC}
 \mathcal{F}^{+C} -\frac{i}{2}\delta^+_{A,B}[ \mathcal{W}, \mathcal{F}
 ], \nonumber\\
 \nabla^+_\mu \nabla^+_\nu \mathcal{F} &=&
  i[\mathcal{V}^+_A,\mathcal{V}^+]+ \frac{i}{32}\Gamma^+_{ABC}
  [\mathcal{V}^{+B},\mathcal{V}^{+C} ]+2\mathcal{F}^-_{\underline{A}}
  +\frac{i}{2}\delta_{\mu\nu}[ \mathcal{F},\mathcal{W} ], \nonumber\\
\nabla^+ \overline{\nabla}^+_\mu \mathcal{W} &=&-i\nabla_{\underline{\mu}}
 \mathcal{V}^+ +i\nabla^{\underline{\nu}} \mathcal{V}^+_{\mu\nu},
 \nonumber\\
\overline{\nabla}^+_\mu \nabla^+_A \mathcal{W} &=&-i
 \delta^+_{A,\mu\nu}\nabla^{\underline{\nu}}\mathcal{V}^+ -i
 \nabla_{\underline{\mu}} \mathcal{V}^+_A+
 \frac{i}{4}\Gamma^+_{AB\mu\nu}\overline{\nabla}^{^+\nu}
 \mathcal{V}^{+B}, \nonumber\\
\nabla^+_\mu \overline{\nabla}^+_\nu \mathcal{W} &=&
 i\delta_{\mu\nu}[\mathcal{V}^+,\mathcal{F}]
 +i[\mathcal{V}^+_{\mu\nu},\mathcal{F}], \nonumber\\
 \overline{\nabla}^+_A \nabla^+_B \mathcal{W}&=& -i\delta^+_{A,B}
  [\mathcal{V}^+,\mathcal{W}]+\frac{i}{4}\Gamma^+_{ABC}[\mathcal{V}^{+C},
  \mathcal{W}], \nonumber\\
\nabla^+\nabla^+_A \mathcal{W}&=&\frac{i}{32}\Gamma^+_{ABC}[
 \mathcal{V}^{+B},  \mathcal{V}^{+C}]+i[ \mathcal{V}^+_A ,
 \mathcal{V}^+] -2 \mathcal{F}^+_{\underline{A}}, \nonumber\\
\nabla^+_A\nabla^+_B \mathcal{W}&=&\frac{i}{4}\Gamma^+_{ABC}
 [\mathcal{V}^{+C},\mathcal{V}^+]+i[\mathcal{V}^+_A,\mathcal{V}^+_B ]
 +\frac{1}{2}\Gamma^+_{ABC}\mathcal{F}^{+C} +\frac{i}{2}\delta^+_{A,B}
 [ \mathcal{W}, \mathcal{F} ], \nonumber\\
 \overline{\nabla}^+_\mu \overline{\nabla}^+_\nu \mathcal{W} &=& -
  i[\mathcal{V}^+_A,\mathcal{V}^+]+ \frac{i}{32}\Gamma^+_{ABC}
  [\mathcal{V}^{+B},\mathcal{V}^{+C} ] +2\mathcal{F}^-_{\underline{A}}
  -\frac{i}{2}\delta_{\mu\nu}[ \mathcal{F}, \mathcal{W} ].
\end{eqnarray}
We define the component fields of these superfields $\mathcal{F}$,
$\mathcal{W}$, $\mathcal{V}^+$ and $\mathcal{V}^+_A$  as the following
forms: 
\begin{align}
\mathcal{F} |&= \phi,  &\overline{\nabla} \mathcal{F}| &= \lambda,&
 \nabla_\mu \mathcal{F}| &= C_\mu,& \overline{\nabla} _A \mathcal{F}| &=
 \lambda^+_A,&    \mathcal{V}| &= -v , \nonumber\\
\mathcal{W}| &=-\overline{\phi},  & \nabla \mathcal{W}| &= \chi,&
 \overline{\nabla}_\mu \mathcal{W} |&= \psi_\mu,& \nabla _A \mathcal{W}| &=
 \chi^+_A,&    \mathcal{V}^+_A | &= v^+_A . \label{N=4 components}
\end{align}
We can derive $N=4$ twisted supertransformations by using the above
equations. The $N=4$ twisted supertransformations strictly correspond
with the twisted ones for Donaldson-Witten theory coupled to the
hypermultiplet at on-shell level in \ref{ch:N=4 symmetry}.

\section{Euclidean $N=4$ super Yang-Mills Action}
We derive an ordinary $N=4$ super Yang-Mills theory by untwisting the
twisted $N=4$ super Yang-Mills theory of the Vafa-Witten type in this
section. We construct Dirac-K\"ahler fermions from the tensor fermions
$\{\lambda, \lambda^+_A, \psi_\mu, \chi, \chi^+_A, C_\mu\}$ appearing in
the twisted $N=4$ super Yang-Mills theory. 
We, however, cannot construct one Dirac-K\"ahler fermion form these
fermion\footnote{It should be noted that in the case of Marcus type an
ordinary Dirac-K\"ahler fermion without projections is constructed by
fermions $\{\tilde{\psi}_\mu, \tilde{\eta}, \chi^-_A, \chi, \chi^+_A,
C_\mu\}$ in four dimensions \cite{KM}.}. The reason is that the
Dirac-K\"ahler fermion needs to a self-dual tensor field and a
anti-self-dual tensor field but this action does not contain self-dual
one. Thus we define two Dirac-K\"ahler fermions with the chiral
projection $(P_+)_{IJ}$ where $I,J=1, \cdots, 4$: 

\begin{eqnarray}
\Psi^{i N=2}_\alpha &\equiv& \psi_{\alpha J}(P_+)_{J I} =\frac{1}{2}
 \{(\lambda + \psi^\mu \gamma_\mu + \frac{1}{4}
 \lambda^{+\mu\nu}\gamma_{\mu\nu})P_+ \}_{\alpha I}, \nonumber\\
\chi^{i N=2}_\alpha &\equiv& \chi_{\alpha J}(P_+)_{J I} =\frac{1}{2}
 \{(\chi + C^\mu \gamma_\mu + \frac{1}{4}\chi^{+\mu\nu}
 \gamma_{\mu\nu})P_+ \}_{\alpha I}. \label{eq:DK fermions}
\end{eqnarray}
$I=3$ and $I=4$ components are vanished in these fermions because of the
chiral projection $P_+$. It means that these fermions are the $N$=2
fermions.  

The action (\ref{eq: on-shell DW coupled to hypermultiplet}) can be
represented by the Dirac-K\"ahler fermions (\ref{eq:DK
fermions}). 
\begin{eqnarray}
\mathcal{S} &=& -\frac{1}{2}\int d^4x \mbox{Tr}\nonumber \\ 
& \times  &
 \Big{(}-\mathcal{D}^\mu v  \mathcal{D}_\mu v
-\frac{1}{4}\mathcal{D}^\mu v^A\mathcal{D}_\mu v_A
+\phi D^\mu D_\mu \overline{\phi} + (F^- _{\mu\nu})^2 \nonumber\\
& &
-i\overline{\Psi}^{i N=2}\gamma^\mu
 \mathcal{D}_\mu \Psi^{i N=2}
-i\overline{\chi}^{i N=2}\gamma^\mu \mathcal{D}_\mu \chi^{i N=2}
\nonumber\\
& &
-2i\overline{\phi}\  \overline{\Psi}^{i N=2}P_+ \Psi^{i
 N=2}  +2i\phi \  \overline{\Psi}^{i N=2}P_- \Psi^{i N=2} \nonumber\\
& &
+2i \phi \  \overline{\chi}^{i N=2}P_+ \chi^{i N=2}
-2i\overline{\phi} \  \overline{\chi}^{i N=2}P_- \chi^{i N=2}
\nonumber\\
& &
-2i v\  \overline{\Psi}^{i N=2}P_+ \chi^{i N=2}
+2iv \  \overline{\Psi}^{i N=2}P_- \chi^{i N=2} \nonumber\\
& &
-2iv\  \overline{\chi}^{i N=2}P_+ \Psi^{i N=2}
+2iv \  \overline{\chi}^{i N=2}P_- \Psi^{i N=2} \nonumber\\
& &
-2i
 (-\frac{1}{4}\hat{\sigma}_{\mu\nu}v^{+\mu\nu})^{ij}\overline{\Psi}^{i
 N=2}P_+ \chi^{j N=2} +2i
 (-\frac{1}{4}\hat{\sigma}_{\mu\nu}v^{+\mu\nu})^{ij}\overline{\Psi}^{i
 N=2}P_- \chi^{j N=2} \nonumber\\
& &
+2i
 (-\frac{1}{4}\hat{\sigma}_{\mu\nu}v^{+\mu\nu})^{ij}\overline{\chi}^{i
 N=2}P_+ \Psi^{j N=2} -2i
 (-\frac{1}{4}\hat{\sigma}_{\mu\nu}v^{+\mu\nu})^{ij}\overline{\chi}^{i
 N=2}P_- \Psi^{j N=2} \nonumber\\
 & &
+\frac{1}{2}v[\phi,[\overline{\phi},v]]
+\frac{1}{2}v[\overline{\phi},[\phi,v]]
+\frac{1}{8}v^{+A}[\phi,[\overline{\phi},v^+_A]]
+\frac{1}{8}v^{+A}[\overline{\phi},[\phi,v^+_A]] \nonumber\\
& &
 +\frac{1}{4}[\phi,\overline{\phi}]^2
-\frac{1}{32}[v^+_A,v^+_B][v^{+A},v^{+B}]-\frac{1}{4}[v^+_A,v][v^{+A},v]
\Big{)},
\end{eqnarray}
where $\hat{\sigma}^{\mu\nu}=\frac{1}{2}(\sigma^\mu
\overline{\sigma}^\nu-\sigma^\nu \overline{\sigma}^\mu)$,
$\sigma^\mu = (\sigma^1,\sigma^2,\sigma^3,\sigma^4
)$,$\overline{\sigma}^\mu = (-\sigma^1,-\sigma^2,-\sigma^3,\sigma^4 )$,\\
$\sigma^4=i {\bf 1}_{2\times 2}$. We furthermore construct a fermion $\Psi^I_\alpha$
with the internal SU(4) R-symmetry from the $\Psi^{i N=2}_\alpha$ and
$\chi^{i N=2}_\alpha$ with the internal SU(2) R-symmetry. We define the
fermion $\Psi^I_\alpha$ with the SU(4) R-symmetry as follows:
\begin{eqnarray}
\Psi^{I}_\alpha &\equiv& \Psi^{i N=2}_{\alpha}, \hspace{20mm} \{I=1,2\},
 \nonumber\\
\Psi^{I}_\alpha &\equiv& \chi^{i N=2}_{\alpha}, \hspace{20mm} \{I=3,4\}.
\end{eqnarray}
We construct fields $\phi^{IJ}$ and $\tilde{\phi}^{IJ}$ of a {\bf 6}
representation with SU(4) symmetry from the bosonic fields: $\phi$, $\overline{\phi}$, $v$ and $v^+_A$ as follows:
\begin{eqnarray}
\phi^{IJ} &=&
\left(
 \begin{array}{cc|cc}
\overline{\phi}&0 & v+iv_{12}&-v_{13}-iv_{14}\\
0& \overline{\phi} & v_{13}-iv_{14}&  v-iv_{12}\\
\hline
 v-iv_{12}&v_{13}+iv_{14} &-\phi &0 \\
- v_{13}+iv_{14}&  v+iv_{12}&0 &-\phi
\end{array}
\right), \nonumber\\
\tilde{\phi}^{IJ} &=&
\left(
 \begin{array}{cc|cc}
-\phi&0 & -v-iv_{12}& v_{13}+iv_{14}\\
0& -\phi & -v_{13}+iv_{14}&  -v+iv_{12}\\
\hline
- v+iv_{12}&-v_{13}-iv_{14} &\overline{\phi} &0 \\
 v_{13}-iv_{14}&  -v-iv_{12}&0 &\overline{\phi}
\end{array}
\right).
\end{eqnarray}
The $\phi^{IJ}$ and $\tilde{\phi}^{IJ}$ are satisfied with the following
relations:
\begin{align}
\phi^\dagger_{IJ} &=-\phi_{IJ}, & \tilde{\phi}^\dagger_{IJ}
 &=-\tilde{\phi}_{IJ},
\end{align}
\begin{eqnarray}
(C\tilde{\phi})^*_{IJ} = -\frac{1}{2}\epsilon_{IJKL}(C\phi)^{KL},
\end{eqnarray}
where $\phi$, $\overline{\phi}$, $v$ and $v^+_A$ are the anti-hermitian.
We can interpret the $\Psi^{I}_\alpha$ as spinors in the $N=4$ SYM theory
since the Lorentz symmetry in twisted theory separate the Lorentz
symmetry into the internal R-symmetry. 
We reparametrize component fields as
\begin{align}
\Psi^I &\to i\Psi^I,&  \phi^{IJ} &\to -i\phi^{IJ},& \tilde{\phi}^{IJ} &\to
 -i \tilde{\phi}^{IJ},
\end{align}
and we derive a $N=4$ super Yang-Mills action as follows:
\begin{eqnarray}
\mathcal{S} &=& -\int d^4x \mbox{Tr} \Big{(}
\frac{1}{8} \tilde{\phi}^{IJ}\mathcal{D}^\mu \mathcal{D}_{\mu}\phi^{JI}
+\frac{i}{2} \overline{\Psi}^I \gamma^\mu \mathcal{D}_\mu \Psi^{I}
+\frac{1}{2}(F^-_{\mu\nu})^2 \nonumber\\
& &\hspace{12mm}+\overline{\Psi}^I P_+ \Psi^J
 \phi^{IJ}+\overline{\Psi}^I P_- \Psi^J \tilde{\phi}^{IJ}
-\frac{1}{64} [\tilde{\phi}^{IJ},\tilde{\phi}^{LM}][\phi^{JI},\phi^{ML}]
\Big{)}.
\end{eqnarray}

\section{Conclusions and Discussions}
We have constructed the Vafa-Witten theory by using $N=2$ or $N=4$ twisted
superspace formalism based on the Dirac-K\"ahler mechanism.
We summarize the Dirac-K\"ahler mechanism in this paragraph.
The Dirac-K\"ahler mechanism gives the way to identify the $N=4$
extended SUSY suffix $\{i\}$ as the Lorentz spinor suffix $\{\alpha\}$,
i.e. , diagonal subgroup $SO(4)\otimes SO(4)_I$\cite{KKM}.
The $N$=4 fermions based on the Dirac-K\"ahler mechanism consist
of two scalar, two vector, a self-dual tensor and an anti-self-dual
tensor. These fields just correspond to the fields which appear in
the Marcus's theory. Since Vafa-Witten theory possesses two anti-self-dual
fields, we cannot construct one Dirac-K\"ahler fermion. We can, therefore,
construct two Dirac-K\"ahler fermion on which is imposed the chiral
projection $(P_+)_{ij}$ with respect to the R-symmetry. 

We proposed the $N=2$ twisted superspace formalism of the twisted
hypermultiplet with the central charge based on the Dirac-K\"ahler
twist. We then introduced the bosonic superfields with a scalar and an
anti-self-dual tensor indices, $\mathcal{V}^+$ and $\mathcal{V}^+_A$
respectively, while we introduced the bosonic superfields
$\mathcal{V}_\mu$ with a vector index in the previous paper. 
The theory given by using the superfield with the vector index is as a result
Marcus's theory \cite{KM}.
In the scalar and the tensor superfields case we have derived the
off-shell action with the auxiliary fields $K^+$ and $K^+_A$ by imposing
the constraints on the superfields from the R-symmetry. 
We have also construct the twisted gauge covariant action with the
Wess-Zumino gauge and have derived the off-shell Donaldson-Witten theory
coupled to the covariantized hypermultiplet.
Integrating out these auxiliary fields $K^+$, $K^+_A$ and $\phi^+_A$, we
then found that the Donaldson-Witten theory coupled to the
hypermultiplet possessed the discrete symmetries. 
From these discrete symmetries we can derive other fermionic symmetries 
whose generators are $\{\overline{s}^+,\overline{s}^+_\mu,
\overline{s}^+_A\}$.  
Thus the symmetries of the theory is enhanced to the $N=4$
twisted SUSY without a central charge. 
The ghost number of 
$\{\overline{s}^+, \overline{s}^+_\mu, \overline{s}^+_A\}$ are
$\{-1, +1, -1\}$, while the ghost number of $\{s^+, s^+_\mu, s^+_A\}$
are $\{+1, -1, +1\}$, where two scalar supercharges $s^+$ and
$\overline{s}^+$ are identified with the BRST charge and the anti-BRST
charge, respectively. We claim that this twist is identified with
Vafa-Witten twist. We explain this more precisely.
We can immediately represent the Dirac-K\"ahler
fermions by using these twisted supercharges as follows:
\begin{eqnarray}
Q^{1}_{\alpha i}&\equiv&\left\{\left(s^{+}+\gamma^\mu s^{+}_\mu +\frac{1}{4}
\gamma^{\mu\nu}s^{+}_{\mu\nu}\right)P_{+}\right\}_{\alpha i}, \nonumber\\
Q^{2}_{\alpha i}&\equiv&\left\{\left(\overline{s}^{+}+\gamma^\mu
\overline{s}^{+}_\mu +\frac{1}{4}\gamma^{\mu\nu}
\overline{s}^{+}_{\mu\nu}\right)P_+\right\}_{\alpha
i}, \label{supercharge of Vafa-Witten type}
\end{eqnarray}
where we define $\{Q^1_{\alpha i}, Q^2_{\alpha i}\}$ as $Q^{\mathcal{A}}_{\alpha
i}$. We can then represent $\{i, \mathcal{A} \}\in\{1,2\}$ as the
suffixes of the $SU(2)_i$ and $SU(2)_{\mathcal{A}}$ group, respectively. 
In this theory, the internal R-symmetry group is
$SU(2)_i\otimes SU(2)_{\mathcal{A}} \simeq SO(4)_I$.  We have shown
that the untwisted theory possesses $N$=4 SUSY with $SO(4)_I$
R-symmetry. As the manner of the twist are 
well known in the papers \cite{Yam, VW}, the Vafa-Witten's twist is
given by taking the diagonal sum of  $SU(2)_i$ of the R-symmetry group
$SU(2)_i\otimes SU(2)_{\mathcal{A}} $ and $SU(2)_L$
of the rotation group $SU(2)_L\otimes SU(2)_R$.
In two component spinor notation, $Q_{\alpha i}^A$ is divided into
$\mathsf{Q}_\alpha^{{\mathcal{A}}i}$ and $\overline{\mathsf{Q}}^{\mathcal{A}}_{\dot{\alpha} i}$
where $\overline{\mathsf{Q}}^{\mathcal{A}}_{\dot{\alpha} i}$ is independent of
$\mathsf{Q}_\alpha^{{\mathcal{A}}i}$ in Euclidean spacetime.
We then identify internal $SU(2)$ index $i$ with $SU(2)_L$ index
$\beta$ : $\mathsf{Q}_\alpha^{{\mathcal{A}}i} \to
\mathsf{Q}_\alpha^{{\mathcal{A}} \beta},\quad \overline{\mathsf{Q}}^{\mathcal{A}}_{\dot{\alpha}
i} \to \overline{\mathsf{Q}}^{\mathcal{A}}_{\dot{\alpha} \beta}  $.
The supercharge $\mathsf{Q}_\alpha^{{\mathcal{A}} \beta}$ is expressed by the
two scalar charges  and the two anti-self-dual charges. On the other hand
the supercharge $ \overline{\mathsf{Q}}^{\mathcal{A}}_{\dot{\alpha} \beta} $
is expressed by two vector charges. We then found that the
Dirac-K\"ahler twist of $Q_{\alpha i}^{\mathcal{A}}$ with chiral projection is
equivalent to the Vafa-Witten's twist after rewriting the tensor
representations with the spinor representations. 
We found that this action corresponded to the Vafa-Witten theory because
of possessing the same ghost number of charges and the same algebras.
Untwisting the theory, we have immediately derived the $N=4$ super
Yang-Mills theory by using the Dirac-K\"ahler mechanism.

One of the main aim is to establish the off-shell theory in the $N=4$
twisted superspace. 
We have tried to construct the $N=4$ twisted superconnection formalism of the
twisted vector multiplet without a central charge.
We then introduced the bosonic curvature superfields $\{\mathcal{F},
\mathcal{W}\}$ and $\{\mathcal{V}^+, \mathcal{V}^+_A\}$ corresponding to
the Donaldson-Witten theory and the twisted hypermultiplet,
respectively. We have imposed the special constraints on the
anti-commutation relations of the fermionic supercovariant derivatives
$\{\nabla^+, \nabla^+_\mu, \nabla^+_A, \overline{\nabla}^+,
\overline{\nabla}^+_\mu, \overline{\nabla}^+_A\}$ and constructed the
$N=4$ twisted super Yang-Mills theory.
Unfortunately, we have naturally derived the equations of motion from
the Jacobi identities of the supercovariant derivatives and derived the
same twisted supertransformations as the Donaldson-Witten theory coupled
to the twisted hypermultiplet at the on-shell level.

Some topological twists were classified by the way of taking diagonal
sum for two global symmetries, i.e., the Lorentz symmetry and the internal
R-symmetry in Euclidean flat spacetime \cite{W1,Yam,Mac,VW}.
We are particularly interested in the Dirac-K\"ahler twist which is one
of the representations of the topological twists because the
Dirac-K\"ahler fermions correspond to the Kogut-Susskind
or the staggered fermions on the lattice. In the recent years the twisted
supersymmetric theory is applied to the lattice theory.
We have constructed the Vafa-Witten theory by using the Dirac-K\"ahler
mechanism in this paper. We may therefore construct this theory on
the lattice.

\vspace{1cm}
\noindent
\textbf{\Large Acknowledgements}\\
We would like to thank Prof. N. Kawamoto for the collaboration at an
earlier stage and instructive comments.
We would like to thank I. Kanamori, K. Nagata and  J. Saito for
fruitful discussions and comments.

\appendix
\setcounter{equation}{0}
\def\thesection{Appendix \Alph{section}}
\renewcommand{\theequation}{A.\arabic{equation}}
\section{}
\label{ch:N=4 symmetry}
we show the full list of the on-shell $N$=4 twisted SUSY transformations
and the R-transformations.
\[
\begin{array}{|c||c|c|c|}
\hline
 &\mbox{gh}\sharp& s^+ & s^+_\mu \\
\hline
v &0& \frac{1}{2}\lambda &-\frac{1}{2} \psi_\mu\\
v^+_B &0& -\frac{1}{2}\lambda ^+_B & \frac{1}{2}\delta^+_{B\mu\nu}\psi^\nu\\
\lambda &1&-i[\phi,v]& -i\mathcal{D}_\mu v
 +i \mathcal{D}^\nu v^+_{\mu\nu}\\
\lambda^+_B &1 &i[\phi,v^+_B]& 
i\delta^+_{B,\mu\nu}\mathcal{D}^\nu v
+i\mathcal{D}_\mu v^+_B - \frac{i}{4}
\Gamma^+_{BC\mu\nu}\mathcal{D}^\nu v^{+C}\\
\psi_\nu &-1  &i\mathcal{D}_\nu v
 + i \mathcal{D}^\rho v^+_{\nu\rho}&
-i \delta_{\mu\nu} [\overline{\phi},v]
+i[\overline{\phi},v^+_{\mu\nu}] \\
\phi& 2 & 0 & C_\mu \\
\overline{\phi} &-2& -\chi & 0 \\
\chi &-1& \frac{i}{2}[\phi, \overline{\phi}] & i \mathcal{D}_\mu
 \overline{\phi}  \\
\chi^+_B& -1&\frac{i}{32}\Gamma^+_{BCD}[v^{+C},v^{+D}]
&-i\delta^+_{B,\mu\nu}\mathcal{D}^\nu  \overline{\phi}\\
& & -i[v^+_B,v]-2F^+_B&  \\
C_\nu & 1 &-i\mathcal{D}_\nu \phi &
 \frac{i}{32}\Gamma^+_{AB\mu\nu}[v^{+A},v^{+B}]-i[v^+_{\mu\nu},v] \\
 & & &+2F^-_{\mu\nu}
-\frac{i}{2}\delta_{\mu\nu}[\phi, \overline{\phi} ] \\
A_\nu & 0&-\frac{i}{2}C_\nu &-\frac{i}{2}(\delta_{\mu\nu} \chi -\chi^+_{\mu\nu}) \\
\hline
\end{array}
\]

\[
\begin{array}{|c||c|c|}
\hline
 & s^+_A & R^+_A\\
\hline
v&\frac{1}{2}\lambda^+_A & -\frac{i}{2}v^+_A \\
v^+_B
 &\frac{1}{2}\delta^+_{A,B}\lambda+\frac{1}{8}\Gamma^+_{ABC}\lambda^{+C}
 & \frac{i}{2}\delta^+_{A,B}v -\frac{i}{8}\Gamma^+_{ABC}v^{+C} \\ 
\lambda & -i[\phi,v^+_A]&0\\
\lambda^+_B &-i\delta^+_{A,B}[\phi,v]
+\frac{i}{4}\Gamma^+_{ABC}[\phi,v^{+C}]     & 0 \\
\psi_\nu & - i\delta^+_{A,\nu\rho}\mathcal{D}^\rho
 v+ i\mathcal{D}_\nu v^+_A 
 -\frac{i}{4}\Gamma^+_{AB\nu\rho}\mathcal{D}^\rho v^{+B} 
&0 \\
\phi & 0 & 0\\
\overline{\phi} &- \chi^+_A &0 \\
\chi & -\frac{i}{32}\Gamma^+_{ABC}[v^{+B},v^{+C}]+i[v^+_A,v]+2F^+_A & -\frac{i}{2}\chi^+_A \\
\chi^+_B &
 i[v^+_A,v^+_B]-\frac{i}{4}\Gamma^+_{ABC}
 [v^{+C},v]+\frac{1}{2}
 \Gamma^+_{ABC}F^{+C}
 & \frac{i}{2}\delta^+_{A,B}\chi -\frac{i}{8}\Gamma^+_{ABC}\chi^{+C} \\
 &   + \frac{i}{2} \delta^+_{A,B} [\phi,\overline{\phi}] &  \\
C_\nu &i\delta^+_{A,\nu\rho}\mathcal{D}^\rho \phi&-\frac{i}{2}\delta^+_{A,\nu\rho}C^\rho\\
A_\nu & -\frac{i}{2}\delta^+_{A,\nu\rho} C^\rho&0\\
\hline
\end{array}
\]

\[
\begin{array}{|c||c|c|}
\hline
 & \overline{s}^+ &  \overline{s}^+_\mu \\
\hline
v & \frac{1}{2}\chi &-\frac{1}{2} C_\mu\\
v^+_B & \frac{1}{2}\chi^+_B &- \frac{1}{2}\delta^+_{B\mu\nu}C^\nu\\
\lambda & -\frac{i}{2}[\phi, \overline{\phi}] & -i \mathcal{D}_\mu
 \phi  \\
\lambda^+_B &\frac{i}{32}\Gamma^+_{BCD}[v^{+C},v^{+D}]+i[v^+_B,v]-2F^+_B
&i\delta^+_{B,\mu\nu}\mathcal{D}^\nu \phi \\
\psi_\nu & i\mathcal{D}_\nu \overline{\phi} &
 \frac{i}{32}\Gamma^+_{AB\mu\nu}[v^{+A},v^{+B}]+i[v^+_{\mu\nu},v] 
+2F^-_{\mu\nu} \\
 & & +\frac{i}{2}\delta_{\mu\nu}[\phi, \overline{\phi} ] \\
\phi & \lambda & 0 \\
\overline{\phi} & 0 & -\psi_\mu \\
\chi &i[\overline{\phi},v]& -i\mathcal{D}_\mu v
 -i \mathcal{D}^\nu v^+_{\mu\nu}\\
\chi^+_B & i[\overline{\phi},v^+_B]& 
i\delta^+_{B,\mu\nu}\mathcal{D}^\nu v
-i\mathcal{D}_\mu v^+_B + \frac{i}{4}
\Gamma^+_{BC\mu\nu}\mathcal{D}^\nu v^{+C}\\
C_\nu &  i\mathcal{D}_\nu v
 - i \mathcal{D}^\rho v^+_{\nu\rho}&
i \delta_{\mu\nu} [\phi,v]
+i[\phi,v^+_{\mu\nu}] \\
A_\nu & -\frac{i}{2}\psi_\nu &-\frac{i}{2}(\delta_{\mu\nu} \lambda -\lambda^+_{\mu\nu}) \\
\hline
\end{array}
\]

\[
\begin{array}{|c||c|c|}
\hline
 & \overline{s}^+_A & \overline{R}^+_A\\
\hline
v&\frac{1}{2}\chi^+_A & \frac{i}{2}v^+_A \\
v^+_B
 &-\frac{1}{2}\delta^+_{A,B}\chi -\frac{1}{8}\Gamma^+_{ABC}\chi^{+C}
 & -\frac{i}{2}\delta^+_{A,B}v-\frac{i}{8}\Gamma^+_{ABC}v^{+C}\\ 
\lambda & -\frac{i}{32}\Gamma^+_{ABC}[v^{+B},v^{+C}]-i[v^+_A,v]+2F^+_A &
-\frac{i}{2}\lambda^+_A \\
\lambda^+_B &
 i[v^+_A,v^+_B]+\frac{i}{4}\Gamma^+_{ABC}
 [v^{+C},v]+\frac{1}{2}
 \Gamma^+_{ABC}F^{+C}
 &\frac{i}{2}\delta^+_{A,B}\lambda
 -\frac{i}{8}\Gamma^+_{ABC}\lambda^{+C}\\
& -\frac{i}{2}\delta^+_{A,B}[\phi,\overline{\phi}]&  \\
\psi_\nu &- i\delta^+_{A,\nu\rho}\mathcal{D}^\rho
 \overline{\phi}&-\frac{i}{2}\delta^+_{A,\mu\nu}\psi^\nu\\
\phi & \lambda^+_A &0 \\
\overline{\phi} & 0 & 0\\
\chi & -i[ \overline{\phi},v^+_A]&0\\
\chi^+_B &i\delta^+_{A,B}[ \overline{\phi},v]
+\frac{i}{4}\Gamma^+_{ABC}[ \overline{\phi},v^{+C}]     & 0 \\
C_\nu & - i\delta^+_{A,\nu\rho}\mathcal{D}^\rho v
- i\mathcal{D}_\nu v^+_A 
+\frac{i}{4}\Gamma^+_{AB\nu\rho}\mathcal{D}^\rho v^{+B} 
&0 \\
A_\nu & -\frac{i}{2}\delta^+_{A,\nu\rho}  \psi^\rho&0\\
\hline
\end{array}
\]
The $N$=4 twisted supercharge $\{s^+,s^+_\mu,s^+_A,\overline{s}^+
,\overline{s}^+_\mu,\overline{s}^+_A \}$ satisfy the following
commutation relations up to the gauge transformation at on-shell level.
\begin{align*}
\{s^+,s^+_\mu \}\varphi&=
 -i(\partial_\mu\varphi+\delta_{\mbox{g}(-\omega_\mu)}\varphi),  &
 \{s^+_\mu,s^+_A\}\varphi&=i\delta^+_{A,\mu\nu}(\partial^\nu\varphi
 +\delta_{\mbox{g}(-\omega^\nu)}\varphi)\\ 
\{\overline{s}^+,\overline{s}^+_\mu \}\varphi&=
 -i(\partial_\mu\varphi+\delta_{\mbox{g}(-\omega_\mu)}\varphi),  &
 \{\overline{s}^+_\mu,\overline{s}^+_A\}\varphi&=i\delta^+_{A,\mu\nu}(\partial^\nu\varphi
 +\delta_{\mbox{g}(-\omega^\nu)}\varphi)
\end{align*}
\begin{align*}
\{s^+,s^+\}\varphi&=\delta_{\mbox{g}(-\phi)} \varphi,
& \{s^+_\mu,s^+_\nu\} \varphi &= \delta_{\mu\nu} 
 \delta_{\mbox{g}(\overline{\phi})} \varphi,
& \{s^+_A,s^+_B\} \varphi &= \delta^+_{A,B}  \delta_{\mbox{g}(-\phi)}
 \varphi, \\
\{\overline{s}^+,\overline{s}^+\}\varphi&=\delta_{\mbox{g}(\overline{\phi})} \varphi,
& \{\overline{s}^+_\mu,\overline{s}^+_\nu\} \varphi &= \delta_{\mu\nu} 
 \delta_{\mbox{g}(-\phi)} \varphi,
& \{\overline{s}^+_A,\overline{s}^+_B\} \varphi &= \delta^+_{A,B}  \delta_{\mbox{g}(\overline{\phi})}
 \varphi, \\
\{s^+,\overline{s}^+_\mu\}\varphi &= 0,
&\{s^+,\overline{s}^+_A\}\varphi &= \delta_{\mbox{g}(-v^+_A)} \varphi,
&\{s^+_A ,\overline{s}^+_\mu\}\varphi &=0 \\
\{\overline{s}^+,s^+_\mu\}\varphi &= 0,
&\{\overline{s}^+,s^+_A\}\varphi &= \delta_{\mbox{g}(v^+_A)} \varphi,
&\{\overline{s}^+_A ,s^+_\mu\}\varphi &=0 
\end{align*}
\begin{align*}
\{s^+_A,\overline{s}^+_B\}\varphi &= \delta^+_{A,B}
 \delta_{\mbox{g}(v)}\varphi -\frac{1}{4}\Gamma^+_{ABC}
 \delta_{\mbox{g}(v^{+C})}\varphi, 
&\{s^+,\overline{s}^+\}\varphi &= \delta_{\mbox{g}(v)} \varphi,\\
\{s^+_\mu,\overline{s}^+_\nu\}\varphi&= -\delta_{\mu\nu}
 \delta_{\mbox{g}(v)}\varphi+   \delta_{\mbox{g}(v^+_{\mu\nu})}\varphi,
\end{align*}
\begin{align*}
\{s^+,s^+_A\}\varphi&= 0, & \{\overline{s}^+,\overline{s}^+_A\}\varphi&= 0,
\end{align*}
where $ \delta_{\mbox{g}(\epsilon)}\varphi \equiv
i[\epsilon,\varphi]$ and $\delta_{\mbox{g}(\epsilon)} A_\mu = D_\mu
\epsilon$ .

\setcounter{equation}{0}
\renewcommand{\theequation}{B.\arabic{equation}}
\section{}
\label{ch:definition of gamma matrices}
In this appendix we define Euclidean  four dimensional $\gamma$-matrices:
\begin{eqnarray}
\{\gamma_\mu, \gamma_\nu \} = 2\delta_{\mu\nu}, \qquad 
\gamma^{\mu\dagger} = \gamma^\mu,
\end{eqnarray}
where $\gamma_\mu$  satisfies the Clifford algebra.
 We use the following notations:
\begin{eqnarray}
\gamma_{\mu\nu}\equiv \frac{1}{2}[\gamma_\mu, \gamma_\nu],\qquad
\tilde{\gamma}_\mu\equiv \gamma_\mu\gamma_5.
\end{eqnarray}
$\gamma^{\mu}$ can be constructed from the 2$\times$2 matrices
$\sigma^{\mu}$ and $\bar{\sigma}^{\mu}$ in the following way:
\begin{eqnarray}
\gamma^\mu =
\left(\begin{array}{cc}
0 & i\sigma^\mu \\
i\bar{\sigma}^\mu & 0
\end{array}\right),
\end{eqnarray}
where $\sigma^\mu = (\sigma^1, \sigma^2, \sigma^3, \sigma^4)$ and 
$\bar{\sigma}^\mu = (-\sigma^1, -\sigma^2, -\sigma^3, \sigma^4)$.
$\sigma^{i}$ are Pauli matrices for $i\in \{1,2,3\}$ and $\sigma^4 = i 
{\bf 1}_{2\times 2}$.  A matrix $B$ and a charge conjugation matrix $C$
are defined as follows:
\begin{align}
\gamma_{\mu} &= \eta B^{-1}\gamma_{\mu}^* B ,& B^* B &= \epsilon\bf{1},
  \nonumber\\
\gamma_\mu &= \eta C^{-1}\gamma_\mu ^T C,& C^T &= \epsilon C,
\end{align}
where $(\eta,\epsilon)=(\pm1,-1) $.

In a four dimensional Euclidean space Majorana fermions do not exist
because the factor $\epsilon$ should be equal to $-1$ \cite{Kugo-T}. A Majorana spinor satisfies
the following condition,
\begin{eqnarray}
\psi^* = B\psi,
\label{eq:def majorana}
\end{eqnarray}
which leads
\begin{eqnarray}
\psi = B^*\psi = B^* B \psi.
\end{eqnarray}
Thus the existence of Majorana fermion requires $B^*B=1$.
This condition can not be taken in four dimensional Euclidean space.
We can, however, take a SU(2)$\simeq$USp(2) Majorana fermion and
a USp(4) Majorana fermion which satisfy the following condition,
respectively,  
\begin{eqnarray}
\psi^{i*} &=& \epsilon^{ij} B\psi^j, 
\label{eq:majorana} \\
\psi^{l*} &=& C^{lm} B\psi^m, 
\label{eq:def USp4majorana}
\end{eqnarray}
where $i,j\in \{1,2\},\quad  l,m \in\{ 1,2,3,4\}$ and these fermions
correspond to the fermions which appear in $N$=2 and $N$=4 supersymmetric
theory, respectively. 
In this paper we choose $(\eta,\epsilon)=(1,-1)$ and $C=B=-\gamma_1 \gamma_3$.

\section{}
\label{ch:algebra}
In this appendix we explain the $N$=4 SUSY algebra with USp(4) Majorana
condition in Euclidean spacetime. $N$=4 SUSY algebra is  
\begin{eqnarray}
\{Q_{\alpha I},\overline{Q}_{J \beta }\} = 2 {\bf 1}_{IJ}
(\gamma^\mu)_{\alpha\beta} P_\mu. 
\label{eq:N=2}
\end{eqnarray} 
We can also represent the
four components supercharge as the following form,
\begin{eqnarray}
Q_{\alpha I}&=& 
 \left(
    \begin{array}{cc}
      \mathsf{Q}_{\alpha} {}^i & i \tilde{\mathsf{Q}}_{\alpha\dot{i}} \\
      i\overline{\mathsf{Q}}^{\dot{\alpha}i} &
       \overline{\tilde{\mathsf{Q}}}^{\dot{\alpha}} {}_{\dot{i}}  
    \end{array}
\right),
\end{eqnarray}

\begin{eqnarray}
{\bf 1}_{IJ} = 
\left(\begin{array}{cc}
\delta_i {}^j & 0 \\
0 & \delta_{\dot{i} } {}^{\dot{j}}
\end{array}\right),\qquad
(\gamma^\mu )_{\alpha\beta} = 
\left(\begin{array}{cc}
0 & i (\sigma^\mu)_{\alpha\dot{\beta}} \\
i(\overline{\sigma}^\mu )^{\dot{\alpha}\beta} & 0
\end{array}\right).
\end{eqnarray}
The USp(4) Majorana condition (\ref{eq:def USp4majorana}) is the
following form, 
\begin{eqnarray}
(Q_{\alpha I} )^*&=& C_{\alpha\beta} Q_{\beta J} C^{-1}_{JI}, \\
 \left(
    \begin{array}{cc}
      \mathsf{Q}_{\alpha} {}^i & i \tilde{\mathsf{Q}}_{\alpha\dot{i}} \\
      i\overline{\mathsf{Q}}^{\dot{\alpha}i} &
       \overline{\tilde{\mathsf{Q}}}^{\dot{\alpha}} {}_{\dot{i}}  
    \end{array}
\right)^*&=&  \left(
\begin{array}{cc}
     -\mathsf{Q}^{\alpha} {}_i & i \tilde{\mathsf{Q}}^{\alpha\dot{i}} \\
      i\overline{\mathsf{Q}}_{\dot{\alpha}i} &
       -\overline{\tilde{\mathsf{Q}}}_{\dot{\alpha}} {}^{\dot{i}}  
    \end{array}\right).
\end{eqnarray} 
Using the USp(4) Majorana condition, we define a
$\overline{Q}_{I\alpha}$ as:
\begin{eqnarray}
\overline{Q}_{I\alpha} = (Q_{\alpha I})^\dagger = 
\left(\begin{array}{cc}
(\mathsf{Q}_\alpha {}^i)^* & (i\overline{\mathsf{Q}}^{\dot{\alpha}i})^* \\
(i\tilde{\mathsf{Q}}_{\dot{\alpha} \dot{i} })^*  &
 (\overline{\tilde{\mathsf{Q}}}^{\dot{\alpha}} {}_{\dot{i}} )^* 
\end{array}\right)=
\left(\begin{array}{cc}
-\mathsf{Q}^\alpha {}_i & i\overline{\mathsf{Q}}_{\dot{\alpha} i} \\
i\tilde{\mathsf{Q}}^{\alpha\dot{i}} & -\overline{\tilde{\mathsf{Q}}}_{\alpha} {}^{\dot{i}}
\end{array}\right),
\end{eqnarray}
We can then describe the algebra (\ref{eq:N=2}) with
respect to theses two-component supercharges,
\begin{align}
\{ \mathsf{Q}_{\alpha} {}^i, \overline{\mathsf{Q}}_{\dot{\alpha}j }\} &=2\delta^i {}_j
 (\sigma^\mu)_{\alpha\dot{\alpha}} P_\mu,&
 \{\tilde{\mathsf{Q}}_{\alpha}{}^i, \overline{\tilde{\mathsf{Q}}}_{\dot{\alpha} j}\}
 =2\delta^i {}_j
 (\sigma^\mu)_{\alpha\dot{\alpha}} P_\mu,
\end{align}
where supercharges with upper and lower indices are related through the
$\epsilon$-tensor: $\mathsf{Q}^\alpha =
\epsilon^{\alpha\beta}\mathsf{Q}_\beta$,\ \ $\mathsf{Q}_\alpha =-
\epsilon_{\alpha\beta}\mathsf{Q}^\beta$.

In two component spinor notation, the Dirac-K\"ahler twist give the
following relations, 
\begin{eqnarray}
\mathsf{Q}_\alpha {}^i \to \mathsf{Q}_\alpha {}^\beta= s^+
 \delta_{\alpha} {}^\beta -\frac{1}{4}s^+_{\mu\nu}
 (\sigma^{\mu\nu})_{\alpha} {}^\beta ,\qquad
 \overline{\mathsf{Q}}^{\dot{\alpha}i} \to
 \overline{\mathsf{Q}}^{\dot{\alpha}\beta} =
 s^+_\mu(\overline{\sigma}^\mu)^{\dot{\alpha}\beta}, \nonumber \\
\overline{\tilde{\mathsf{Q}}}^{\dot{\alpha}} {}_{\dot{i}} \to
 \overline{\tilde{\mathsf{Q}}}^{\dot{\alpha}} {}_{\dot{\beta}}= s^-
 \delta^{\dot{\alpha}} {}_{\dot{\beta}} -\frac{1}{4}s^-_{\mu\nu}
 (\sigma^{\mu\nu})^{\dot{\alpha}} {}_{\dot{\beta}} ,\qquad
 \tilde{\mathsf{Q}}_{\alpha\dot{i}} \to
 \tilde{\mathsf{Q}}_{\alpha\dot{\beta}} =
 s^-_\mu( \sigma^\mu)_{\alpha\dot{\beta}} \nonumber.
\end{eqnarray}
On the other hand, Vafa-Witten twist is the following relations,
\begin{eqnarray}
\mathsf{Q}^{\mathcal{A}}_\alpha {}^i \to \mathsf{Q}^{\mathcal{A}}_\alpha {}^\beta= s^{{\mathcal{A}}+}
 \delta_{\alpha} {}^\beta -\frac{1}{4}s^{{\mathcal{A}}+}_{\mu\nu}
 (\sigma^{\mu\nu})_{\alpha} {}^\beta ,\qquad
 \overline{\mathsf{Q}}^{{\mathcal{A}}\dot{\alpha}i} \to
 \overline{\mathsf{Q}}^{{\mathcal{A}}\dot{\alpha}\beta} =
 s^{{\mathcal{A}}+}_\mu(\overline{\sigma}^\mu)^{\dot{\alpha}\beta},
\end{eqnarray}
where ${\mathcal{A}}\in \{1,2\}$ and $\{s^{1+},s^{1+}_\mu,
s^{1+}_{\mu\nu}, s^{2+},s^{2+}_\mu, s^{2+}_{\mu\nu} \} 
=\{ s^{+},s^{+}_\mu, s^{+}_{\mu\nu}, \overline{s}^{+}, \overline{s}^{+}_\mu,  \overline{s}^{+}_{\mu\nu}\}$.

\setcounter{equation}{0}
\renewcommand{\theequation}{D.\arabic{equation}}
\section{}
\label{ch:notations}
In this appendix we give the definition of $\delta^\pm_{A, B}$ and 
$\Gamma^\pm_{ABC}$. The suffix $A$ is the second rank tensor
which denotes the suffix $\mu\nu$: $\mu, \nu \in \{ 1, \cdots,
4$ \}. The definition of $\delta^\pm_{A,B}$ is 
\begin{eqnarray}
\delta^\pm_{A,B} = \delta^{\pm}_{\mu\nu, \rho\sigma}
= \delta_{\mu\rho}\delta_{\nu\sigma}
 -\delta_{\mu\sigma}\delta_{\nu\rho} \mp \epsilon_{\mu\nu\rho\sigma},
\end{eqnarray}
where $\delta^{\pm A,B}\delta^{\mp}_{~A,B}=0$.
(Anti-)self-dual tensors $\chi^{\pm A}$ satisfy 
\begin{eqnarray}
\chi^{\pm A} = \frac{1}{4}\delta^{\pm A,B}\chi_{B},
\end{eqnarray}
where $\chi_{B}=\chi^+_{B} +\chi^-_{B}$. 

Variants of the definition of $\Gamma_{ABC}$ which stand for the third
anti-symmetric tensor for $ ABC $ is 
\begin{eqnarray}
\Gamma^{\pm \mu\alpha, \nu\beta, \rho\gamma} &=& \delta^{\alpha\nu}
\delta^{\beta\rho}\delta^{\gamma\mu} +\delta^{\mu\nu}\delta^{\beta\gamma}
\delta^{\rho\alpha} + \delta^{\alpha\beta}\delta^{\nu\gamma}\delta^{\rho\mu}
+\delta^{\mu\beta}\delta^{\nu\rho}\delta^{\gamma\alpha} \nonumber\\
&& -(\delta^{\alpha\nu}\delta^{\beta\gamma}\delta^{\rho\mu} +\delta^{\mu\nu}
\delta^{\beta\rho}\delta^{\gamma\alpha} +\delta^{\alpha\beta}\delta^{\nu\rho}
\delta^{\gamma\mu} +\delta^{\mu\beta}\delta^{\nu\gamma}\delta^{\rho\alpha})
\nonumber \\
&& \mp\epsilon^{\mu\alpha\beta\gamma}\delta^{\beta\gamma} \mp
\epsilon^{\mu\alpha\nu\rho}\delta^{\beta\gamma} \pm 
\epsilon^{\mu\alpha\nu\gamma}\delta^{\beta\rho} \pm
\epsilon^{\mu\alpha\beta\rho}\delta^{\nu\gamma} \nonumber \\
&=& \delta^{\pm\mu\alpha,\nu\rho}\delta^{\beta\gamma}
 +\delta^{\pm\mu\alpha, 
\beta\gamma}\delta^{\nu\rho} -\delta^{\pm\mu\alpha,\nu\gamma}
\delta^{\beta\rho} -\delta^{\pm\mu\alpha,\beta\rho}\delta^{\nu\gamma}, \\
\Gamma^{\pm AB\rho\sigma} &=& 
\frac{1}{2}(\delta^{\pm A,\nu\rho}\delta^{\pm B},_{\nu}^{~\sigma}
-\delta^{\pm A, \nu\sigma}\delta^{\pm B},_{\nu}^{~\sigma}), \\
\Gamma^{\pm ABC} &=& \frac{1}{4}\delta^{\pm A,\nu\rho}
\delta^{\pm B},_{\nu}^{~\sigma}\delta^{\pm C},_{\rho\sigma}.
\end{eqnarray}

\end{document}